\documentclass[11pt,a4paper,oneside]{report}

\usepackage[utf8]{inputenc}
\usepackage[T1]{fontenc}
\usepackage{lmodern}
\usepackage[a4paper,left=1.4in,right=1in,top=1.2in,bottom=1.2in]{geometry}
\usepackage{setspace}
\usepackage{amsmath,amssymb,amsfonts,amsthm}
\usepackage{bm}
\usepackage{graphicx}
\usepackage[table]{xcolor}
\usepackage{colortbl}
\usepackage{booktabs}
\usepackage{multirow}
\usepackage{array}
\usepackage{tabularx}
\usepackage{longtable}
\usepackage{caption}
\usepackage{subcaption}
\usepackage{float}
\usepackage{enumitem}
\usepackage{pifont}
\newfloat{algorithm}{tp}{loa}
\floatname{algorithm}{Algorithm}
\newcommand{\algrule}{\vspace{2pt}\hrule\vspace{3pt}}
\newenvironment{algsteps}{\begin{list}{\arabic{algline}:}{\usecounter{algline}%
  \setlength{\leftmargin}{1.8em}\setlength{\itemsep}{1pt}\setlength{\parsep}{0pt}}}{\end{list}}
\newcounter{algline}
\usepackage{url}
\usepackage{microtype}
\usepackage{tikz}
\usetikzlibrary{arrows.meta,positioning,shapes.geometric,calc,fit,backgrounds,decorations.pathreplacing}
\usepackage{pgfplots}
\pgfplotsset{compat=1.17}
\usepackage{fancyhdr}
\usepackage{titlesec}
\usepackage[hidelinks,breaklinks]{hyperref}
\hypersetup{colorlinks=true,linkcolor=black,citecolor=black,urlcolor=blue}

\definecolor{iiitblue}{RGB}{20,63,120}
\definecolor{softgray}{RGB}{240,242,245}
\definecolor{accent}{RGB}{176,42,55}
\definecolor{deepgreen}{RGB}{20,110,80}

\titleformat{\chapter}[display]
  {\normalfont\huge\bfseries\color{iiitblue}}
  {\filright\Large\color{accent}CHAPTER \thechapter}{10pt}{\Huge}
\titlespacing*{\chapter}{0pt}{-20pt}{28pt}

\theoremstyle{definition}
\newtheorem{definition}{Definition}[chapter]

\newtheorem{property}{Property}[chapter]
\theoremstyle{plain}

\theoremstyle{remark}

\newcommand{\keybox}[2]{%
  \begin{center}
  \begin{tikzpicture}
  \node[draw=iiitblue,fill=softgray,rounded corners,line width=0.9pt,
        inner sep=8pt,text width=0.92\linewidth] {%
    \textbf{\color{iiitblue}#1}\;\; #2};
  \end{tikzpicture}
  \end{center}}

\newcommand{\real}{\mathbb{R}}
\newcommand{\Enc}{\phi}
\newcommand{\Dset}{\mathcal{D}}
\newcommand{\Loss}{\mathcal{L}}

\newcommand{\decker}{\textsc{Decker}}
\newcommand{\hear}{\textsc{Hear}}
\newcommand{\evflow}{\textsc{EvFlow-CTSSM}}
\newcommand{\lfcan}{\textsc{LF-CAN}}
\newcommand{\asca}{ASCA}

\newcommand{\topk}{top-$k$}

\begin{document}

\thispagestyle{empty}
\begin{center}
\vspace*{1.2cm}

{\LARGE\bfseries\color{iiitblue}
Domain-Invariant and Language-Model-Assisted\\[0.35cm]
Acoustic Side-Channel Attacks on Keyboards\par}

\vspace{2.2cm}
{\large by\par}
\vspace{0.5cm}
{\Large\bfseries Bikrant Bikram Pratap Maurya \\\ [MT24116]\par}

\vspace{2.0cm}
{\large Under the Supervision of\par}
\vspace{0.3cm}
{\large\bfseries Dr.\ Arun Balaji Buduru\par}

\vspace{2.0cm}
\IfFileExists{logo.jpg}{\includegraphics[width=3cm]{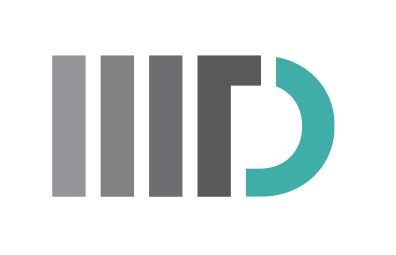}}{%
  \begin{tikzpicture}
  \node[circle,draw=iiitblue,line width=1.2pt,minimum size=2.1cm,
        align=center,text=iiitblue,font=\bfseries] {IIIT\\D};
  \end{tikzpicture}}

\vspace{1.0cm}
{\large\bfseries Computer Science and Engineering\par}
{\large\bfseries Indraprastha Institute of Information Technology Delhi\par}
\vspace{0.4cm}
{\large August, 2026\par}
\end{center}
\clearpage

\thispagestyle{empty}
\begin{center}
\vspace*{0.4cm}
{\small \textcopyright\ Indraprastha Institute of Information Technology (IIITD), New Delhi, 2026\par}

\vspace{1.6cm}
{\LARGE\bfseries\color{iiitblue}
Domain-Invariant and Language-Model-Assisted\\[0.35cm]
Acoustic Side-Channel Attacks on Keyboards\par}

\vspace{1.8cm}
{\large by\par}
\vspace{0.4cm}
{\Large\bfseries Bikrant Bikram Pratap Maurya \\\ [MT24116]\par}

\vspace{1.8cm}
{\large Submitted\par}
\vspace{0.3cm}
{\large in partial fulfillment of the requirements for the degree of\par}
\vspace{0.4cm}
{\Large\bfseries Master of Technology \par}
\vspace{0.4cm}
{\large to\par}

\vspace{1.0cm}
{\large\bfseries Computer Science and Engineering\par}
{\large\bfseries Indraprastha Institute of Information Technology Delhi\par}
\vspace{0.4cm}
{\large August, 2026\par}
\end{center}
\clearpage

\pagenumbering{roman}
\chapter*{Certificate}
\addcontentsline{toc}{chapter}{Certificate}
\thispagestyle{plain}

\vspace{0.6cm}
\noindent This is to certify that the thesis titled
\textbf{``Domain-Invariant and Language-Model-Assisted Acoustic Side-Channel Attacks on
Keyboards''} being submitted by \textbf{Bikrant Bikram Pratap Maurya} to the Indraprastha
Institute of Information Technology Delhi, for the award of the Master of Technology, is an
original research work carried out by him under my supervision. In my opinion, the thesis has
reached the standards fulfilling the requirements of the regulations relating to the degree.

\vspace{0.4cm}
\noindent The results contained in this thesis have not been submitted in part or full to any
other university or institute for the award of any degree/diploma.

\vspace{2.6cm}
\noindent
\begin{minipage}[t]{0.48\textwidth}
August, 2026
\end{minipage}
\hfill
\begin{minipage}[t]{0.48\textwidth}
\raggedright
\textbf{Dr.\ Arun Balaji Buduru}\\
Department of Computer Science and Engineering\\
Indraprastha Institute of Information Technology Delhi\\
New Delhi 110\,020
\end{minipage}

\clearpage

\chapter*{Declaration Regarding Use of AI in Thesis}
\addcontentsline{toc}{chapter}{Acknowledgements}
\thispagestyle{plain}

I acknowledge that I am fully responsible for the entire content of my thesis, including any
sections assisted by online tools, including Artificial Intelligence-based tools. I accept full
accountability for any violations of ethical standards in publications arising from the use of
such tools.

\vspace{0.8cm}
\hfill\textit{Bikrant Bikram Pratap Maurya [MT24116]}

\clearpage

\chapter*{ACKNOWLEDGEMENTS}
\addcontentsline{toc}{chapter}{Acknowledgements}
\thispagestyle{plain}

I am deeply grateful to the Department of Computer Science \& Engineering and to the
Infosys Centre for Artificial Intelligence at IIIT-Delhi. Their provision of
computational infrastructure, GPU resources, and a stimulating research environment made
this work possible.

My foremost thanks go to my advisor, \textbf{Dr.\ Arun Balaji Buduru}, whose guidance
shaped every stage of this thesis. His insistence on rigour, his patience during the long
data-collection effort behind the \hear{} dataset, and his encouragement to pursue an
ambitious open-world formulation of acoustic side-channel attacks were invaluable. He
taught me to treat a security threat model not as a checkbox but as the backbone of an
honest evaluation, and that lesson runs through the whole of this document.

I am indebted to my collaborators \textbf{Nitin Choudhury},
\textbf{Orchid Chetia Phukan} and \textbf{Bhavin Vinodbhai Kuwar} . The \decker{} framework and the VoIP representation and
transduction studies reported in this thesis grew directly out of our joint work, our late-night
recording sessions, and our many arguments about whether a keyboard was leaking its own
identity or the key that was pressed. Their contributions to the underlying publications
are gratefully acknowledged.


Finally, I thank my family and friends for their unwavering support through the demanding
months of coursework, experimentation, and writing. Their belief in me was the quiet
constant behind this work.

\vspace{0.8cm}
\hfill\textit{Bikrant Bikram Pratap Maurya}

\clearpage

\chapter*{ABSTRACT}
\addcontentsline{toc}{chapter}{Abstract}
\thispagestyle{plain}


Acoustic side-channel attacks (\asca{}) recover typed text from the sound of keystrokes, using
only a commodity microphone and no access to the victim's device. Prior work establishes that
such attacks succeed under idealized laboratory conditions---a single keyboard, a fixed
microphone, a quiet room, and training and testing on the same device---but leaves open the
question that matters for security: whether keystroke inference survives the open world of
unseen keyboards, diverse users, noisy environments, and compressed transmission channels. The
answer developed here is that it does, provided the acoustic model is made explicitly
domain-invariant and its noisy predictions are corrected by a language model.

The central contribution is \decker{} (Domain-invariant Embedding for Cross-Keyboard Extraction
and Recognition), a keystroke-inference framework that unifies Keyboard Signature Normalization,
domain-adversarial disentanglement, supervised cross-keyboard contrastive alignment, and Acoustic
Style Randomization to produce embeddings that are discriminative of key identity yet invariant to
keyboard, user, and environment. \decker{} narrows the cross-keyboard generalization gap from
roughly forty accuracy points---the collapse suffered by strong unimodal and multimodal baselines
that encode device identity rather than key identity---to under eighteen. A sentence-level attack
then applies constrained-beam-search language-model decoding over the top-$k$ acoustic candidates,
substantially boosting recovery of coherent text and quantifying the risk for password classes of
varying entropy.

Two further contributions support and extend this result. \hear{}, the dataset underpinning every
experiment, comprises keystroke acoustics from fifty-three participants on thirty-seven laptop
keyboards, captured through an external microphone, the on-device microphone, and a Voice-over-IP
(VoIP) channel, and annotated with demographic and user-awareness metadata. Building on \hear{}'s
VoIP recordings, a companion study of speech pretrained models under codec distortion shows that a
Kolmogorov--Arnold Network adaptation head recovers much of the accuracy lost to compression, and
\evflow{}, a continuous-time transduction model, canonicalizes distorted VoIP embeddings to make
fully segmentation-free attacks viable. Together these results argue that domain-invariant
representation learning is a central requirement for any credible evaluation of \asca{}, and that
modern language models materially amplify its real-world impact---findings that motivate the
hardware-, system-, and model-level defenses discussed in the closing chapters.

\vspace{.8cm}
\noindent\textbf{KEYWORDS:} \quad Acoustic Side-Channel Attack; ASCA; Keystroke Inference;
Domain Invariance; Cross-Keyboard Generalization; ECAPA-TDNN; Domain-Adversarial Learning;
Supervised Contrastive Learning; Acoustic Style Randomization; Speech Pretrained Models;
Kolmogorov--Arnold Networks; Voice-over-IP; Large Language Models; Constrained Beam Search;
Security and Privacy.

\clearpage

\tableofcontents
\clearpage
\addcontentsline{toc}{chapter}{List of Figures}
\listoffigures
\clearpage
\addcontentsline{toc}{chapter}{List of Tables}
\listoftables
\clearpage
\chapter*{ABBREVIATIONS}
\addcontentsline{toc}{chapter}{Abbreviations}
\thispagestyle{plain}

\begin{longtable}{p{2.6cm} p{11cm}}
\textbf{ASCA} & Acoustic Side-Channel Attack \\
\textbf{HEAR} & The keystroke-acoustics dataset introduced in this thesis (the single dataset used throughout) \\
\textbf{DECKER} & Domain-invariant Embedding for Cross-Keyboard Extraction and Recognition \\
\textbf{EvFlow-CTSSM} & Event-Triggered Continuous-Time State-Space Model with Latent Flow Canonicalization \\
\textbf{LF-CAN} & Latent Rectified Flow Canonicalization \\
\textbf{CT-SSM} & Continuous-Time State-Space Model \\
\textbf{CTC} & Connectionist Temporal Classification \\
\textbf{CER} & Character Error Rate \\
\textbf{AED} & Acoustic Event Detection \\
\textbf{KSN} & Keyboard Signature Normalization \\
\textbf{ASR} & Acoustic Style Randomization (in the modeling chapters) \\
\textbf{ASR} & Automatic Speech Recognition (only where explicitly noted) \\
\textbf{ECAPA} & Emphasized Channel Attention, Propagation and Aggregation \\
\textbf{TDNN} & Time-Delay Neural Network \\
\textbf{GRL} & Gradient Reversal Layer \\
\textbf{DANN} & Domain-Adversarial Neural Network \\
\textbf{SupCon} & Supervised Contrastive (loss) \\
\textbf{PTM} & Pretrained Model \\
\textbf{SSL} & Self-Supervised Learning \\
\textbf{KAN} & Kolmogorov--Arnold Network \\
\textbf{FCN} & Fully Connected Network \\
\textbf{CNN} & Convolutional Neural Network \\
\textbf{LLM} & Large Language Model \\
\textbf{VoIP} & Voice over Internet Protocol \\
\textbf{AGC} & Automatic Gain Control \\
\textbf{AEC} & Acoustic Echo Cancellation \\
\textbf{NLoS} & Non-Line-of-Sight \\
\textbf{LoS} & Line-of-Sight \\
\textbf{MFCC} & Mel-Frequency Cepstral Coefficient \\
\textbf{FFT} & Fast Fourier Transform \\
\textbf{SNR} & Signal-to-Noise Ratio \\
\textbf{IIR} & Infinite Impulse Response \\
\textbf{t-SNE} & t-distributed Stochastic Neighbor Embedding \\
\textbf{ID} & In-Domain \\
\textbf{OOD} & Out-of-Domain \\
\textbf{mF1} & Macro-averaged F1 score \\
\textbf{PII} & Personally Identifiable Information \\
\textbf{TASLP} & (IEEE/ACM) Transactions on Audio, Speech, and Language Processing \\
\textbf{TIFS} & (IEEE) Transactions on Information Forensics and Security \\
\end{longtable}

\clearpage

\chapter*{NOTATION}
\addcontentsline{toc}{chapter}{Notation}
\thispagestyle{plain}

\begin{longtable}{p{3.0cm} p{10.6cm}}
$x_i \in \real^{T}$ & Raw keystroke waveform of $T$ samples for the $i$-th event \\
$y_i \in \{1,\dots,|K|\}$ & Ground-truth key label of the $i$-th keystroke \\
$d_i \in \{1,\dots,|\Dset|\}$ & Keyboard (domain) identity of the $i$-th keystroke \\
$\Enc(\cdot)$ & Acoustic encoder mapping a waveform to an embedding \\
$z_i = \Enc(x_i) \in \real^{E}$ & Embedding of the $i$-th keystroke, dimension $E$ \\
$h_{d_i}$ & Device-specific acoustic impulse response of keyboard $d_i$ \\
$s_i$ & Latent (device-free) keystroke source signal \\
$g_\theta(\cdot)$ & Learnable inverse filter (KSN module) \\
$G_\theta(f)$ & Frequency-domain response of KSN \\
$C_k(\cdot)$ & Key classifier head \\
$C_d(\cdot)$ & Domain (keyboard) classifier head \\
$\text{GRL}_\lambda(\cdot)$ & Gradient reversal layer with strength $\lambda$ \\
$P(i),\,N(i)$ & Positive / negative index sets for contrastive alignment \\
$\tau$ & Contrastive temperature \\
$\lambda_{dom},\lambda_{con}$ & Loss weights for domain and contrastive terms \\
$\Loss_{key}$ & Key-classification (cross-entropy) loss \\
$\Loss_{dom}$ & Domain-adversarial loss \\
$\Loss_{supcon}$ & Supervised contrastive loss \\
$\mathcal{C}_t$ & Top-$k$ candidate key set at timestep $t$ \\
$P_{\text{LM}}(Y)$ & Language-model prior over token sequence $Y$ \\
$\lambda_{\text{LM}}$ & Language-model fusion weight in beam search \\
$\Phi_q(\cdot),\psi_{q,p}(\cdot)$ & Learnable univariate functions of a KAN layer \\
$\mathbb{E}[\cdot]$ & Expectation \\
$\|\cdot\|_1,\|\cdot\|_2$ & $\ell_1$ and $\ell_2$ norms \\
$*$ & Convolution operator \\
\end{longtable}

\clearpage

\clearpage
\pagenumbering{arabic}

\chapter{Introduction}
\label{ch:intro}

\section{General Overview}

An acoustic side-channel attack (\asca{}) on a keyboard rests on a deceptively simple
physical fact: pressing a key produces a short, characteristic sound, and that sound can
serve as a fingerprint for the key on a given device. If an adversary can capture those
sounds and map them back to keys, then a microphone---an object now embedded in nearly
every phone, laptop, earbud, and conferencing appliance---becomes an instrument for
recovering everything a victim types, from search queries and private messages to
passwords and PINs. Unlike a keylogger, this attack requires no software on the victim's
machine; unlike shoulder-surfing, it requires no line of sight; and unlike network
interception, it is indifferent to encryption, because it operates on the physical act of
typing rather than on the data after it has left the keyboard.

The threat is not hypothetical. A line of research stretching back two decades has shown,
across many setups and threat models, that keystroke acoustics carry enough structure to
identify keys with surprising accuracy~\cite{asonov2004,zhuang2009,harrison2023,berger2006,
halevi2012,compagno2017}. Early work relied on hand-crafted spectral features and classical
classifiers; more recent systems apply deep convolutional and transformer architectures to
spectrograms and raw waveforms, and the best of them report per-key accuracies that would
be alarming if they held in practice. The last two years have added a further multiplier.
Large language models (LLMs) encode exactly the linguistic regularities---character
$n$-grams, word morphology, syntax, and common password patterns---that let an attacker
correct a noisy stream of key predictions into coherent text~\cite{ayati2025,radford2019gpt2,
raffel2020t5}. A single imperfect acoustic model, followed by a language model, can plausibly
reconstruct a sentence even when a third of its raw key predictions are wrong.

Yet almost all of this evidence was gathered under narrow, idealized conditions:
a single keyboard, a fixed microphone geometry, a quiet room, a homogeneous set of
typists, and training and testing on the very same device. These are precisely the
conditions least like the settings in which the attack would actually be launched. In a
café, a library, or an online meeting, the adversary does not control the victim's
hardware, cannot fix the recording geometry, must contend with ambient noise and multipath
reflections, and will encounter keyboards, switch mechanisms, and typing styles never seen
during training. The uncomfortable question this thesis confronts is therefore not
``\emph{can} keystrokes be recovered in the lab?''---that is settled---but ``\emph{does the
attack survive} the open world?''

\section{Setting up the Flow}

This thesis argues, and then demonstrates, that the answer is a qualified yes. The
qualification is important, because the naive answer is no. When we take strong modern
acoustic models---self-supervised speech encoders, spectrogram vision transformers, and
their multimodal fusions---and evaluate them honestly on keyboards they have never heard,
their accuracy collapses, typically by forty absolute percentage points
(Table~\ref{tab:unimodal} in Chapter~\ref{ch:results}). The reason is subtle and is the intellectual core of the
thesis: these models do not learn a representation of \emph{keys}, they learn a
representation of the \emph{keyboard}. The device's impulse response---its switch type, key
travel, chassis resonance, and the geometry between keys and microphone---imprints itself on
every recording, and a model trained to maximize accuracy on one device happily exploits
that imprint. Move to a new device and the imprint changes, so the model's confident
predictions become confident nonsense.

The remedy is to make the acoustic model explicitly \emph{domain-invariant}: to strip away
device-specific colouration while preserving the information that distinguishes one key from
another. This thesis develops such a model, \decker{}, and pairs it with a language-model
decoding stage that corrects the residual key-level errors using linguistic structure. The
combination restores much of the accuracy lost to keyboard shift and, at the sequence
level, recovers coherent text and human-like passwords far more often than the acoustic
model alone. Along the way, the thesis contributes the datasets that make an honest
evaluation possible in the first place, because the field's progress has been throttled less
by modeling ideas than by the absence of realistic, diverse data.

\section{Objectives of the Thesis}

\subsection{The Research Gap}

Three gaps in the \asca{} literature motivate this work.

\paragraph{Data gap.} Existing \asca{} datasets are highly constrained: typically
single-keyboard, single-environment, and demographically narrow, with little variation in
gender, typing background, or laptop usage profile, and almost no coverage of realistic
transmission channels such as Voice-over-IP (VoIP)~\cite{harrison2023,wang2024reflexnoop}. This
makes it impossible to study whether \asca{} generalizes beyond laboratory conditions, and
it obscures the role of user and demographic factors in the attack. No prior dataset jointly
covers multiple devices, multiple users, multiple genders, noisy public environments,
non-line-of-sight capture, continuous natural-language text, and VoIP or smartphone streams.

\paragraph{Modeling gap.} Deep \asca{} models overfit to keyboard- and setup-specific
spectral patterns, and their performance drops sharply on unseen devices, indicating that
they encode device identity rather than pure keystroke content. Multimodal fusion of audio
and spectrogram features adds capacity but does not remove this domain shift, because both
encoders carry the same device colouration. There is no established framework that produces
embeddings which are simultaneously key-discriminative, keyboard-invariant, and consistent
for the same key across devices.

\paragraph{Sequence gap.} Even when per-key predictions are reasonably accurate, local
errors accumulate over long sequences such as passwords and sentences. Most pipelines treat
keys independently and do not systematically exploit linguistic structure to correct errors.
Where language models have been used, they have appeared as loosely coupled filters on top of
keyboard-specific acoustic models, never as a principled decoder over the top-$k$ candidate
distribution of a domain-invariant encoder.

\subsection{Research Questions}

The thesis is organized around six research questions that the datasets and models are
designed to answer.

\begin{description}[leftmargin=1.4cm,style=nextline]
\item[RQ1] Do \asca{} models generalize across keyboards, and if so, under what modeling
conditions?
\item[RQ2] Why do strong unimodal and multimodal baselines fail under keyboard, user, and
environment shift?
\item[RQ3] Can a single training framework produce embeddings that are key-discriminative
yet keyboard-, user-, and environment-invariant?
\item[RQ4] How do modern speech pretrained representations behave under VoIP codec
distortion, and can a more expressive adaptation head recover the lost accuracy?
\item[RQ5] What is the role of large language models in \asca{}, and how much do they raise
the effective threat at the sequence level?
\item[RQ6] What does a realistic, end-to-end \asca{} threat model look like once the
idealized assumptions on hardware, environment, and users are abandoned?
\end{description}

\section{A Short Literature Survey}

The modern study of keyboard \asca{} begins with Asonov and Agrawal, who showed that a
neural network trained on FFT features could identify keys from the sound of typing
alone~\cite{asonov2004}. Zhuang, Zhou, and Tygar removed the need for labelled training data
by combining unsupervised clustering with a language model, foreshadowing the LLM-assisted
decoding that is now central~\cite{zhuang2009}. A decade of work then broadened the attack
to smartphones, touchscreens, VoIP calls, and video conferencing~\cite{compagno2017,
sabra2020zoom,yu2021touch,lu2019keylistener}, while the deep-learning era brought CNNs,
residual networks, and transformer spectrogram models that pushed laboratory accuracy
higher~\cite{harrison2023,martinasek2015}. Most recently, RefleXnoop and Heimdall demonstrated
non-line-of-sight recovery by exploiting multipath reflections, and LLM-assisted spectrogram
correction demonstrated the sequence-level multiplier that language priors
provide~\cite{wang2024reflexnoop,luo2024heimdall,ayati2025}. Chapter~\ref{ch:related} develops
this history in full; the point to carry forward here is that the trajectory has been one of
steadily relaxing assumptions, and that the assumption most in need of relaxing---training and
testing on the same device---is the one this thesis targets.

\section{Motivation: Why Study ASCA Now?}

Three trends make \asca{} a more credible threat today than at any point in its history.
First, high-quality far-field microphones are ubiquitous and cheap; a concealed smartphone
or a compact USB microphone captures enough detail for key classification. Second, deep
architectures for noisy acoustic signals have matured, so an attacker no longer needs
bespoke signal processing to extract fine transient cues. Third, and most consequentially,
capable language models are freely available, and they convert an unreliable per-key
classifier into a reliable sentence and password reconstructor. The convergence of these
trends means that the barrier to a practical attack is now low: a passive adversary with a
commodity microphone and a laptop-class machine, seated at a socially unremarkable distance,
can plausibly recover typed text from a victim using a keyboard the attacker has never seen.
Establishing whether that sentence is true, and quantifying exactly how true, is the purpose
of this thesis.

\section{Contributions}

The thesis makes the following contributions, each developed in a dedicated chapter and each
grounded in the two publications that underlie this work.

\begin{enumerate}[leftmargin=1.5em]
\item \textbf{\hear{}: a realistic \asca{} dataset.} We curate \hear{}, a dataset of
keystroke acoustics from $53$ participants typing on $37$ laptop keyboards, recorded under
three realistic conditions (external microphone, on-device microphone, and VoIP streaming),
and annotated with demographic attributes and user-awareness responses. \hear{} is the first
dataset to jointly cover multi-device, multi-user, multi-gender, noisy-environment,
non-line-of-sight, continuous-text, and VoIP/smartphone-stream factors
(Chapter~\ref{ch:hear}).

\item \textbf{A comprehensive \asca{} benchmark.} On \hear{} we establish a benchmark spanning
conventional features and strong pretrained representations, on raw waveforms and
spectrograms, in unimodal and multimodal settings, and we quantify how far existing
approaches generalize across users, keyboards, and environments
(Chapters~\ref{ch:hear} and~\ref{ch:results}).

\item \textbf{\decker{}: a domain-invariant inference framework.} We introduce \decker{},
which combines Keyboard Signature Normalization, domain-adversarial disentanglement,
supervised cross-keyboard contrastive alignment, and Acoustic Style Randomization to learn
key-discriminative yet keyboard-invariant embeddings (Chapter~\ref{ch:decker}).

\item \textbf{VoIP-channel robustness on \hear{}.} Using the VoIP recordings of \hear{}, we
provide the first systematic evaluation of self-supervised and supervised speech pretrained
models for \asca{} under VoIP codec distortion and propose a Kolmogorov--Arnold Network
adaptation head that consistently outperforms fully connected and convolutional baselines. We
further introduce \evflow{}, a continuous-time transduction model that canonicalizes distorted
VoIP embeddings in the latent space and decodes unsegmented streams, making fully
segmentation-free VoIP attacks viable (Chapter~\ref{ch:keyac}).

\item \textbf{LLM-assisted reconstruction and risk quantification.} We investigate a
sentence-level attack that applies constrained-beam-search LLM decoding to refine noisy
keystroke predictions, and we quantify the resulting risk for natural text and for password
classes of varying entropy (Chapters~\ref{ch:decker} and~\ref{ch:results}).
\end{enumerate}

\section{Organization of the Thesis}

Chapter~\ref{ch:background} develops the physical, algorithmic, and linguistic background:
the acoustics of keystrokes, deep architectures for audio, speech pretrained models,
Kolmogorov--Arnold networks, domain-adaptation theory, and the role of language models.
Chapter~\ref{ch:related} surveys two decades of \asca{} research and positions this work.
Chapter~\ref{ch:threat} specifies the realistic threat model. Chapter~\ref{ch:hear} presents
the \hear{} dataset---the single dataset used throughout the thesis---and Chapter~\ref{ch:keyac}
studies \asca{} over VoIP channels on \hear{}, covering both speech-representation robustness and
the \evflow{} continuous-transduction model. Chapter~\ref{ch:decker} develops the \decker{}
framework in full mathematical detail. Chapter~\ref{ch:results} reports the experimental results.
Chapter~\ref{ch:discussion} discusses implications, defenses, and limitations, and
Chapter~\ref{ch:conclusion} concludes and outlines future work, including the extension path
toward a journal (TASLP) submission. Appendices provide the standardized typing corpus, the
metadata schema, extended result tables, and the transaction-extension plan.

\chapter{Background and Preliminaries}
\label{ch:background}

This chapter assembles the conceptual machinery the rest of the thesis relies on. It moves
from the physics of a keystroke, through the acoustic and deep-learning tools that turn a
recording into a key prediction, to the representation-learning, domain-adaptation, and
language-modeling ideas that make the attack robust and its output coherent. A reader
already fluent in speech representation learning and domain-adversarial training may skim
Sections~\ref{sec:bg-ptm}--\ref{sec:bg-da}, but the acoustics of Section~\ref{sec:bg-acoustics}
and the language-model decoding of Section~\ref{sec:bg-llm} are worth reading closely, because
they justify specific design choices in \decker{}.

\section{The Acoustics of a Keystroke}
\label{sec:bg-acoustics}

Pressing a key is a small mechanical event with a surprisingly rich acoustic signature. The
finger drives the keycap downward; the switch mechanism actuates, whether a scissor mechanism
on a laptop, a rubber dome on a membrane board, or a spring on a mechanical switch; the keycap
bottoms out against the chassis; and, on release, the key returns and a second, softer
transient is produced. Each of these sub-events radiates a short broadband pulse, and the
superposition---shaped by keycap vibration, chassis resonance, air-pressure release, and
reverberation from nearby surfaces---forms the sound that a microphone records.

Two empirical regularities make the attack possible. First, on a fixed device each key emits a
\emph{relatively consistent} pattern across presses, despite variability in typing style, so
that a classifier trained on one user's presses of a key transfers to other presses of the same
key~\cite{asonov2004,zhuang2009}. Second, keys that are spatially adjacent tend to share
acoustic characteristics, because their vibration paths through the chassis overlap, whereas
keys in distinct keyboard regions differ in spectral distribution and decay. This spatial
structure is what lets unsupervised methods bootstrap key clusters without labels.

A convenient and, for this thesis, load-bearing way to model the recording is as a convolution.
Let $s_i$ be the latent, device-independent source signal of the $i$-th keystroke---the sound
the key ``wants'' to make---and let $h_{d_i}$ be the impulse response of device $d_i$, which
lumps together switch mechanics, chassis resonance, microphone frequency response, and room
acoustics. Then the observed waveform is
\begin{equation}
x_i \;=\; h_{d_i} * s_i,
\label{eq:conv-model}
\end{equation}
where $*$ denotes convolution. Equation~\eqref{eq:conv-model} is the single most important
equation in the thesis, because it localizes the generalization problem precisely: the key
identity lives in $s_i$, but every model sees $x_i$, which is $s_i$ smeared through the
device-specific filter $h_{d_i}$. A model that maximizes accuracy on one device learns to read
$s_i$ \emph{through} a fixed $h_{d_i}$; change the device and $h_{d_i}$ changes, and the learned
decoder mismatches. Keyboard Signature Normalization (Chapter~\ref{ch:decker}) is, at its core,
a learnable attempt to invert $h_{d_i}$ and recover an estimate of $s_i$.

\subsection{Multipath Propagation and Environmental Effects}

Keystroke sound does not reach the microphone by a single path. It reflects off the laptop
screen, the desk surface, the user's hands, and surrounding walls, and these reflections
introduce predictable but nonlinear modifications to the spectral envelope and temporal decay of
the signal. The laptop screen, in particular, often acts as a stable secondary reflector,
producing additional peaks in the waveform that can \emph{increase} distinguishability in some
geometries while degrading it in others~\cite{wang2024reflexnoop}. Systems such as
RefleXnoop and Heimdall exploit exactly these multipath cues to recover keystrokes with no direct
line of sight~\cite{wang2024reflexnoop,luo2024heimdall}. The double edge of multipath is central
to the threat model of Chapter~\ref{ch:threat}: reflections are simultaneously an enabler of
non-line-of-sight attacks and a source of device- and room-specific variability that a robust
model must suppress.

\subsection{Segmentation: from a Stream to Keystroke Windows}

Before any classifier runs, a continuous recording must be sliced into per-keystroke windows.
Given a keystroke timestamp $t_{\text{ms}}$, a fixed asymmetric window
\begin{equation}
t_{\text{start}} = t_{\text{ms}} - 60\,\text{ms}, \qquad t_{\text{end}} = t_{\text{ms}} + 200\,\text{ms}
\label{eq:window}
\end{equation}
captures both the impulsive onset and the early reverberant tail, consistent with prior
\asca{} practice. Windows shorter than a minimum energy threshold are discarded as
noise-only, and rapid typing can cause partial overlap between adjacent windows, which is
retained deliberately because it reflects realistic attack conditions. In a fully passive
attack the timestamps are not known and must be estimated from onset detection; the sensitivity
of the whole pipeline to segmentation error is studied in Chapter~\ref{ch:results}.

\section{Signal Preprocessing and Feature Extraction}
\label{sec:bg-features}

Before any learned model runs, a keystroke window is transformed into a representation the model can consume.
Two families dominate, and both appear in this thesis: hand-crafted spectral features and learned front-ends.

\paragraph{Short-time analysis.} A windowed waveform $x[n]$ is analyzed frame by frame with the short-time
Fourier transform (STFT),
\begin{equation}
X[m,\omega] = \sum_{n} x[n]\, w[n-mH]\, e^{-j\omega n},
\end{equation}
where $w[\cdot]$ is an analysis window (Hann, length $25$\,ms) and $H$ is the hop ($10$\,ms). The squared
magnitude $|X[m,\omega]|^2$ is the power spectrogram, whose sharp onset structure is exactly the transient a
keystroke produces.

\paragraph{Mel and log-mel spectrograms.} Human-and machine-relevant frequency resolution is non-uniform, so
the power spectrum is warped onto the mel scale, $\text{mel}(f) = 2595\log_{10}(1 + f/700)$, by a bank of $F$
triangular filters $\{H_k\}$. The log-mel feature is
\begin{equation}
S[m,k] = \log\!\Big(\sum_\omega H_k[\omega]\,|X[m,\omega]|^2 + \epsilon\Big),
\end{equation}
and the resulting $S\in\real^{F\times L}$ is the input to the spectrogram models and to the ECAPA-TDNN encoder of
Section~\ref{sec:bg-ecapa}. Log compression matters here because keystroke energy spans a wide dynamic range
between the impulsive onset and the reverberant tail.

\paragraph{MFCCs.} Applying a discrete cosine transform to the log-mel spectrum and keeping the low-order
coefficients yields mel-frequency cepstral coefficients (MFCCs), $c[m,q] = \sum_k S[m,k]\cos\!\big(q(k-\tfrac12)\pi/F\big)$.
MFCCs decorrelate the filter-bank energies and were the dominant feature in classical \asca{}; they serve here as a
conventional baseline against which learned representations are compared.

\section{Deep Architectures for Acoustic Classification}

\subsection{Convolutional and Transformer Backbones}

Convolutional neural networks (CNNs) remain the workhorse for spectrogram classification: a
stack of learned filters detects local time--frequency patterns, pooling builds translation
tolerance, and depth composes simple detectors into discriminative features~\cite{oshea2015cnn}.
Residual networks add skip connections that ease the training of deep stacks, and hybrid
convolution--attention models such as CoAtNet marry the locality of convolution with the global
receptive field of self-attention, which is well suited to spectrograms where a key's identity
depends on both fine transient structure and coarse spectral shape~\cite{dai2021coatnet}. For raw
waveforms, time-domain encoders avoid the fixed time--frequency trade-off of a spectrogram and can
learn front-ends tuned to the signal; the ECAPA-TDNN encoder, developed for speaker verification,
is a prominent example and is the acoustic backbone of \decker{}.

\subsection{ECAPA-TDNN}
\label{sec:bg-ecapa}

ECAPA-TDNN (Emphasized Channel Attention, Propagation and Aggregation, Time-Delay Neural
Network) extracts frame-level features with multi-scale one-dimensional convolutions,
recalibrates channels with squeeze-and-excitation attention, and aggregates frames with
attentive statistics pooling~\cite{desplanques2020ecapa}. Two properties make it attractive for
\asca{}. First, its multi-scale temporal convolutions capture both the sharp onset and the longer
decay of a keystroke, which live at different time scales. Second, its attentive pooling produces a
single fixed-dimensional utterance-level embedding from a variable-length frame sequence, which is
exactly the interface a key classifier and a contrastive loss want. Formally, for a log-mel input
$S_i \in \real^{F\times L}$ the encoder produces frame features $H_i = \phi(S_i) \in \real^{C\times
L}$, and statistics pooling forms
\begin{equation}
\mu_i = \frac{1}{L}\sum_{t=1}^{L} H_i(:,t), \qquad
\sigma_i = \sqrt{\frac{1}{L}\sum_{t=1}^{L}\big(H_i(:,t)-\mu_i\big)^2},
\end{equation}
so that the pooled vector $[\mu_i;\sigma_i]$ summarizes the sequence and a linear head maps it to
the embedding $z_i$.

\section{Speech Pretrained Models}
\label{sec:bg-ptm}

A pretrained model (PTM) learns general-purpose representations from large unlabelled or
weakly-labelled corpora and is then adapted to a downstream task. Speech PTMs have transformed
audio classification, and the VoIP study of Chapter~\ref{ch:keyac} asks how well they transfer
to the unusual domain of impulsive keystroke sounds. We use six.

\paragraph{Self-supervised backbones.} \textbf{Wav2Vec2} learns contextualized speech
representations by masked prediction of latent units with a contrastive objective~\cite{baevski2020wav2vec2}.
\textbf{HuBERT} uses a BERT-style masked-prediction objective over clustered acoustic units~\cite{hsu2021hubert}.
\textbf{WavLM} extends masked prediction with a denoising objective, jointly modeling acoustic,
phonetic, and speaker cues, which turns out to matter for the noisy, transient signals of \asca{}~\cite{chen2022wavlm}.
\textbf{XLS-R} is a large multilingual Wav2Vec2 variant trained for cross-lingual transfer~\cite{babu2021xlsr}.

\paragraph{Supervised backbones.} \textbf{X-vectors} are embeddings from a TDNN trained for speaker
identification, widely reused for speaker and paralinguistic tasks~\cite{snyder2018xvector,novotny2018xvectors}.
\textbf{Whisper} is an encoder--decoder trained on large-scale weakly-supervised data for multilingual
recognition and translation; we use only its encoder to extract embeddings~\cite{radford2023whisper,kodali2023}.
The relative strengths of these backbones have been mapped by large speech-representation benchmarks such as
SUPERB~\cite{yang2021superb}, SLUE and its second phase~\cite{shon2022slue,shon2023slue2}, and broad
foundation-model evaluations~\cite{yang2024largescale,porjazovski2024}.

For all backbones the standard protocol freezes the encoder and trains a lightweight downstream head on
the extracted embeddings, which isolates the intrinsic quality of the representation from the capacity of
the head---a lightweight adaptation practice established in prior speech work~\cite{phukan2023transforming} and a
distinction that becomes the crux of the VoIP representation findings.

\section{Kolmogorov--Arnold Networks}
\label{sec:bg-kan}

Kolmogorov--Arnold Networks (KANs) are motivated by the Kolmogorov--Arnold representation theorem,
which states that any multivariate continuous function can be written as a finite composition of
univariate continuous functions and addition~\cite{liu2024kan}. Concretely, a function of $n$ variables can
be expressed as
\begin{equation}
f(x_1,\dots,x_n) = \sum_{q} \Phi_q\!\left(\sum_{p}\psi_{q,p}(x_p)\right),
\end{equation}
where the $\psi_{q,p}$ and $\Phi_q$ are learnable univariate functions. A KAN layer operationalizes this
by replacing the fixed scalar weights of a multilayer perceptron with learnable univariate functions on
the edges, typically parameterized as B-splines. A single KAN layer computes
$h_j = \sum_i \phi_{i,j}(x_i)$ with each $\phi_{i,j}$ a spline. The practical consequence is that a KAN can
model nonlinear feature interactions more explicitly than an MLP with fixed activations and scalar weights.
KANs have already proven effective as parameter-efficient adapters for pretrained image models~\cite{cai2025loki}
and for speech enhancement~\cite{li2024grkan}, which is why Chapter~\ref{ch:keyac} adopts a KAN as the adaptation
head over frozen speech representations: when VoIP codecs warp the spectral and temporal statistics of a keystroke
embedding, the mapping from representation to key becomes highly nonlinear, and a spline-based head captures it
where a linear or shallow head cannot.

\section{Domain Adaptation and Invariant Representation Learning}
\label{sec:bg-da}

The generalization crisis of Chapter~\ref{ch:intro} is, in the language of machine learning, a
\emph{domain-shift} problem: training keyboards and test keyboards are drawn from different
distributions, and a model minimizing training risk exploits features that do not transfer. Two
complementary tools address it.

\paragraph{Domain-adversarial training.} A domain-adversarial neural network (DANN) inserts a
\emph{gradient reversal layer} (GRL) between the encoder and an auxiliary domain classifier~\cite{ganin2015dann}.
The GRL is the identity on the forward pass and multiplies the gradient by $-\lambda$ on the backward pass:
\begin{equation}
\text{GRL}_\lambda(z) = z, \qquad \frac{\partial\,\text{GRL}_\lambda}{\partial z} = -\lambda I.
\end{equation}
The domain classifier tries to predict the keyboard from the embedding, but because the gradient is reversed
before it reaches the encoder, the encoder is pushed to make the keyboard \emph{un}predictable while the key
classifier keeps the key predictable. The equilibrium is an embedding that retains key information and
discards keyboard information.

\paragraph{Supervised contrastive alignment.} Adversarial training removes domain information but does not by
itself force same-key embeddings from \emph{different} devices to coincide. Supervised contrastive
learning does~\cite{khosla2020supcon}: it pulls together embeddings that share a label, here the key,
while pushing apart embeddings of different keys, and by drawing positives from different devices it explicitly
enforces cross-keyboard key consistency. Chapter~\ref{ch:decker} shows that adversarial and contrastive terms
are complementary---the former suppresses device identity, the latter aligns key identity---and that both are
needed.

\paragraph{Style randomization.} A third, data-centric tool exposes the model to synthetic device variation.
By perturbing a signal with random filters and spectral warps at training time---the audio analogue of the
SpecAugment idea~\cite{park2019specaugment}---one simulates keyboards that were never recorded, so that the
encoder cannot rely on any single device's colouration. This is the basis of Acoustic Style Randomization.

\section{Language Models in Side-Channel Attacks}
\label{sec:bg-llm}

Typed text is not random. It obeys character-level $n$-gram statistics, word morphology, syntax, and, for
passwords, well-documented human-selection biases. Large language models---from GPT-2 and GPT-3 to
instruction-tuned encoder--decoders~\cite{radford2019gpt2,brown2020gpt3,raffel2020t5}---encode these regularities
and can correct sequences with local errors, a property already exploited in \asca{}-adjacent settings as a
post-processing autocorrect step~\cite{ayati2025,zhuang2009}. The thesis treats language modeling not as an
afterthought but as a decoding stage. Let the acoustic model emit a per-timestep distribution
$P_\theta(y_t\mid x_t)$, restricted to the \topk{} plausible keys $\mathcal{C}_t = \{y_{t,1},\dots,y_{t,k}\}$.
A language model with prior $P_{\text{LM}}(Y)$ then selects the sequence maximizing a fused objective
\begin{equation}
\hat{Y} = \arg\max_{Y\in \mathcal{C}_{1:T}} \left[\sum_{t=1}^{T}\log P_\theta(y_t\mid x_t) + \lambda_{\text{LM}}\log P_{\text{LM}}(Y)\right],
\label{eq:llm-decode}
\end{equation}
implemented with constrained beam search. Equation~\eqref{eq:llm-decode} is deliberately parallel to shallow-fusion
decoding in speech recognition: the acoustic term keeps the reconstruction faithful to what was heard, and the
language term keeps it linguistically plausible, correcting substitution, insertion, and deletion errors that
the acoustic model alone cannot resolve. A theme developed in Chapter~\ref{ch:results} is that the value of the
language term depends on how much human-selection bias has lowered the entropy of the target: it helps natural
text and human-like passwords greatly, and high-entropy random strings almost not at all.

\keybox{Chapter takeaway.}{Keystroke acoustics factor into a device-free source and a device impulse
response (Eq.~\ref{eq:conv-model}); the whole thesis is an argument that robust \asca{} means estimating the
former while discarding the latter, then letting a language model clean up what remains.}

\chapter{Related Work}
\label{ch:related}

Acoustic side-channel analysis has re-emerged as a practical threat because of the convergence of
three trends: the widespread availability of high-quality microphones capable of capturing far-field
audio, rapid progress in deep learning for noisy acoustic signals, and an increasing reliance on mobile
computing in public and semi-public environments. This chapter reviews the physical, algorithmic, and
linguistic strands of the literature, and then positions the present work against the datasets and
systems that precede it.

\section{Foundations of Keyboard Acoustic Emanations}

The field opens with Asonov and Agrawal, who demonstrated that keyboards, telephone keypads, and ATM
pads leak their input acoustically, and that a neural network trained on FFT features of individual
keystrokes could recover keys~\cite{asonov2004}. Their key empirical observation---that each key emits a
distinguishable and repeatable sound on a given device---remains the vulnerability basis for all
subsequent work. Zhuang, Zhou, and Tygar made the attack dramatically more practical by removing the need
for labelled training data: they clustered keystrokes acoustically, then used language-model and
spelling constraints to assign clusters to characters, and finally re-trained a supervised classifier on
the recovered labels~\cite{zhuang2009}. Their pipeline is the intellectual ancestor of the LLM-assisted
decoding studied here, differing mainly in the sophistication of the language prior.

Halevi and Saxena examined how typing style and password randomness affect emanation attacks, tempering
optimism by showing that random passwords and touch-typing degrade recovery relative to hunt-and-peck
entry of natural text~\cite{halevi2012}. Berger, Wool, and Yeredor framed dictionary attacks over
keystroke acoustics as a constraint-satisfaction problem, again foreshadowing the language-driven
decoding that modern LLMs perform far better. Early timing side channels on typed input, such as Song,
Wagner, and Tian's analysis of inter-keystroke timing over SSH, are complementary: they exploit
\emph{when} keys are pressed rather than \emph{how} they sound, and are outside the acoustic scope of this
thesis, but they establish the broader principle that typing leaks through many channels at once~\cite{song2001timing}.

\section{From Signal Processing to Deep Learning}

The modeling side of \asca{} has progressed from hand-crafted features with classical classifiers to
end-to-end deep learning. Initial work relied on temporal and spectral features---FFT coefficients, MFCCs,
and power or energy profiles---combined with statistical classifiers~\cite{asonov2004,zhuang2009}. Foo Kune and
Kim modeled the problem probabilistically, and hidden Markov models were introduced to capture temporal
structure across keystroke sequences~\cite{fookune2010timing}. Martinasek et al.\ applied spectrograms with
neural networks, marking the transition to learned features~\cite{martinasek2015}.

The deep-learning era proper arrives with Harrison, Toreini, and Mehrnezhad, who trained a modern CNN on
laptop keystroke spectrograms and reported high accuracy including over a Zoom (VoIP) channel, demonstrating
that even low-cost microphones and compressed audio retain enough detail for classification~\cite{harrison2023}.
Subsequent systems adopt residual networks, ConvMixer architectures, and audio pretrained models such as
Wav2Vec2, HuBERT, and WavLM on keystroke audio, while waveform-level encoders such as ECAPA-TDNN extract fine
transient cues originally developed for speaker verification~\cite{trockman2022convmixer,desplanques2020ecapa}.
The consistent finding across this body of work is that keystroke acoustics are highly structured and
reproducible even for modern scissor-switch laptop keyboards. The equally consistent and less-advertised
finding, which this thesis foregrounds, is that these models are acutely sensitive to device-specific factors,
so that cross-device generalization is poor.

\section{Relaxing the Line-of-Sight and Geometry Assumptions}

A recent and influential thread relaxes the geometric assumptions of earlier attacks. RefleXnoop shows that
sound reflected from a laptop screen encodes structural information about the keyboard and enables password
snooping on non-line-of-sight (NLoS) laptops, turning a nuisance---multipath---into a signal~\cite{wang2024reflexnoop}.
Heimdall demonstrates keystroke inference in virtual reality by eavesdropping on controller acoustic emanations,
extending \asca{} to entirely new input modalities and confirming that reflective and indirect paths carry
recoverable information~\cite{luo2024heimdall}. These systems substantially broaden the physical attack surface,
but they still largely assume that the attacker can train on data from the target device or a close analogue;
they do not solve the cross-\emph{keyboard} generalization problem, which is orthogonal to the line-of-sight
question.

\section{Language Models as a Sequence-Level Multiplier}

Human-generated keystroke sequences exhibit strong statistical regularities, and language models encode them.
The idea of using linguistic constraints to correct acoustic errors is as old as Zhuang et al.'s spelling-and-
grammar post-processing, but its modern incarnation is far more powerful. Ayati et al.\ make \asca{} on noisy
keyboards viable by using an LLM to perform ``typo'' correction on spectrogram-derived character
predictions, reporting large gains at the sequence level~\cite{ayati2025}. This is representative of how
language models currently enter \asca{}: as loosely coupled filters applied after a keyboard-specific acoustic
model. What is missing---and what this thesis supplies---is the pairing of a genuinely domain-invariant acoustic
encoder, which produces stable \topk{} distributions across keyboards, with a constrained LLM decoder operating
over those distributions under a realistic threat model. The distinction matters because a language model can
only correct errors that are locally recoverable; if the acoustic model's errors are systematic and
device-driven, no amount of linguistic prior will fix them.

\section{Attacks on Other Devices and Modalities}

For completeness we note the breadth of acoustic and related side channels beyond laptop keyboards. Compagno et
al.\ eavesdrop on typing during Voice-over-IP calls~\cite{compagno2017}; Sabra, Maiti, and Jadliwala exploit video
calls for keystroke inference~\cite{sabra2020zoom}; Yu et al.\ and Lu et al.\ infer touchscreen input from
acoustic sensing~\cite{yu2021touch,lu2019keylistener}; and Zhou et al.\ crack Android pattern locks from acoustic
signals~\cite{zhou2018patternlistener}. Das, Borisov, and Caesar fingerprint smartphones through embedded acoustic
components, illustrating that microphones and speakers themselves carry identifying signatures~\cite{das2014fingerprint}.
Wang et al.\ combine keystroke detection across acoustic signals~\cite{wang2016combined}. These works confirm that
acoustic leakage is pervasive across input devices, and several of them---particularly the VoIP and video-call
attacks---directly motivate the VoIP-aware data collection of Chapters~\ref{ch:hear} and~\ref{ch:keyac}.

\section{Surveys and the State of the Field}

The recent survey by Taheritajar, Harris, and Rahaeimehr consolidates two decades of keyboard \asca{} research and,
tellingly, identifies dataset limitations and cross-device generalization as open problems rather than solved
ones~\cite{taheritajar2025survey}. Dissertations and applied studies in the area reach the same
conclusion~\cite{liu2019dissertation,fadake2023beyond}. The field has, in short, thoroughly established that the
attack works in the lab and has begun to relax its geometric assumptions, but it has not established whether the
attack survives realistic device, user, and environment diversity, nor has it built the data needed to find out.

\section{Positioning of This Thesis}

Table~\ref{tab:dataset-compare} summarizes how prior \asca{} datasets compare to \hear{} along the axes that
matter for realistic evaluation. The pattern is clear: earlier datasets vary one or two factors at most, and
none jointly cover device, user, gender, noisy-environment, NLoS, continuous-text, and VoIP/smartphone-stream
diversity. This thesis differs from prior work in three coordinated ways. First, it supplies the data
(\hear{}, including its VoIP subset) to evaluate \asca{} under the diversity real attacks face. Second, it treats
cross-keyboard generalization as the primary evaluation axis and builds a framework (\decker{}) that targets it
directly, rather than reporting same-device accuracy. Third, it integrates language-model decoding as a
principled stage over a domain-invariant encoder's \topk{} outputs, rather than as a bolt-on filter. The result
is the most comprehensive assessment to date of whether, and under what conditions, acoustic keystroke leakage
constitutes a real-world threat.

\begin{table}[t]
\centering
\caption{Comparison of \asca{} datasets across device diversity, user demographics, acoustic conditions, and
recording modalities. \hear{} is the first dataset to jointly include multi-device (MD), multi-user (MU),
multi-gender (MG), noisy public environments (NE), NLoS capture, continuous text, and VoIP/smartphone streams.}
\label{tab:dataset-compare}
\small
\begin{tabularx}{\linewidth}{l c *{7}{>{\centering\arraybackslash}X}}
\toprule
Dataset & Size & MD & MU & MG & NE & NLoS & Cont.\ text & VoIP/phone \\
\midrule
Asonov \& Agrawal (2004)~\cite{asonov2004} & $\sim$2k & \ding{55} & \ding{55} & \ding{55} & \ding{55} & \ding{55} & \ding{55} & \ding{55} \\
Zhuang et al.\ (2009)~\cite{zhuang2009} & $\sim$10k & \ding{55} & \ding{55} & \ding{55} & \ding{55} & \ding{55} & \ding{51} & \ding{55} \\
Harrison et al.\ (2023)~\cite{harrison2023} & 900 & \ding{55} & \ding{55} & \ding{55} & \ding{51} & \ding{55} & \ding{55} & \ding{51} \\
RefleXnoop (CCS'24)~\cite{wang2024reflexnoop} & 5k+ & \ding{55} & \ding{55} & \ding{55} & \ding{51} & \ding{51} & \ding{55} & \ding{55} \\
Heimdall (NDSS'24)~\cite{luo2024heimdall} & 10.4k & \ding{51} & \ding{51} & \ding{55} & \ding{51} & \ding{51} & \ding{55} & \ding{51} \\
\rowcolor{softgray}\hear{} (this thesis) & 10k+ & \ding{51} & \ding{51} & \ding{51} & \ding{51} & \ding{51} & \ding{51} & \ding{51} \\
\bottomrule
\end{tabularx}
\end{table}

\chapter{Threat Model}
\label{ch:threat}

A security evaluation is only as meaningful as the threat model it assumes. Much of the optimism in the
\asca{} literature dissolves under a realistic model, and much of the pessimism dissolves once language-model
decoding is admitted. This chapter fixes a single, conservative threat model that is consistent with modern
non-line-of-sight acoustic attacks such as RefleXnoop and Heimdall, and that governs every experiment in the
thesis. The adversary requires no device compromise, no elevated privileges, and no abnormal physical
proximity; it relies solely on passive audio capture and offline computation, including pretrained language
models for sequence reconstruction.

\section{Attack Scenario}
\label{sec:threat-scenario}

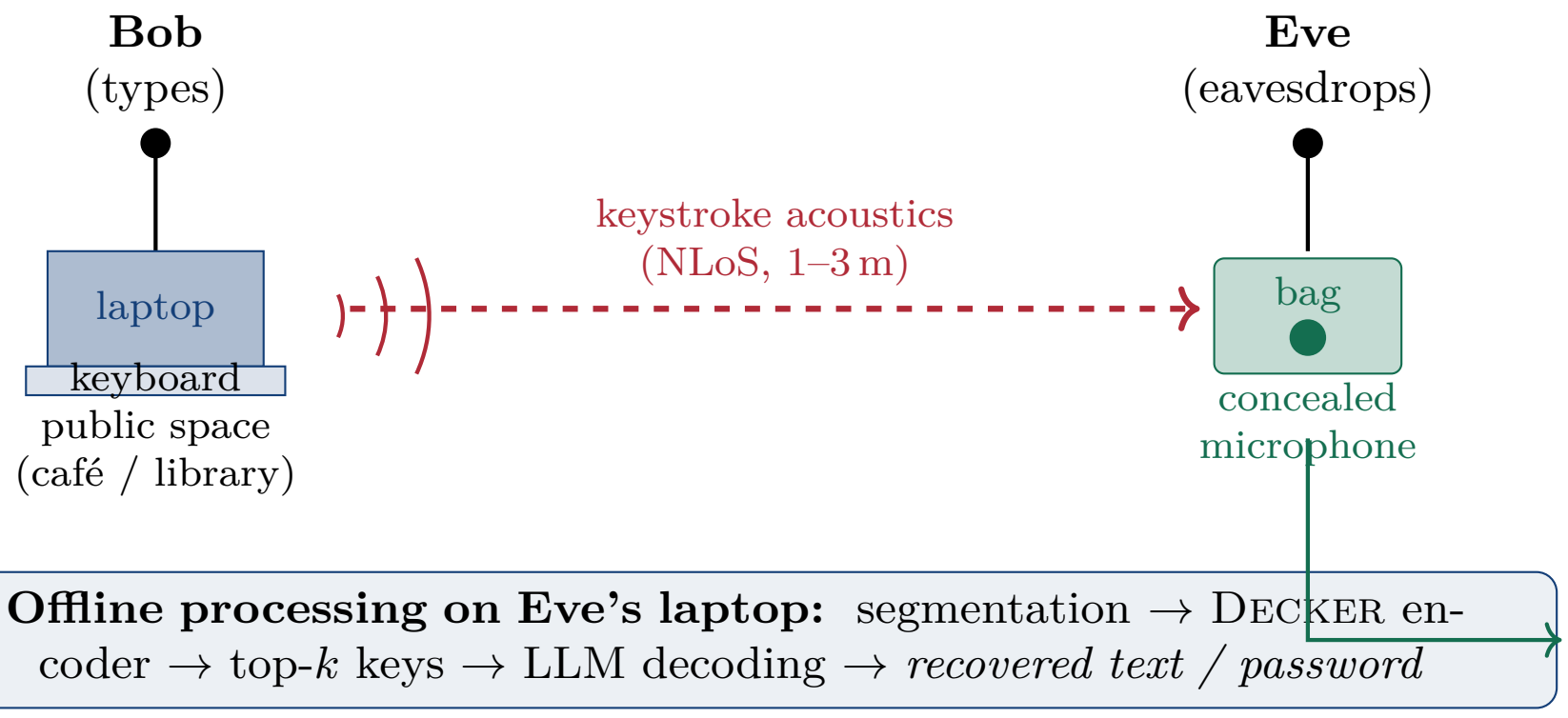
\begin{figure}[t]
\centering
\begin{tikzpicture}[font=\small,
  person/.style={circle,fill=black,minimum size=6pt,inner sep=0pt},
  every node/.style={outer sep=1pt}]

\node[person] (bobhead) at (0,3.1) {};
\draw[thick] (0,3.0)--(0,2.35);
\node[align=center,font=\footnotesize] at (0,3.65) {\textbf{Bob}\\(types)};
\draw[fill=iiitblue!35,draw=iiitblue] (-0.75,1.55) rectangle (0.75,2.35); 
\draw[fill=iiitblue!15,draw=iiitblue] (-0.9,1.35) rectangle (0.9,1.55);  
\node[iiitblue,font=\scriptsize] at (0,1.95) {laptop};
\node[font=\scriptsize] at (0,1.45) {keyboard};
\node[align=center,font=\scriptsize,text=black] at (0,0.95) {public space\\(caf\'e / library)};

\foreach \r in {0.35,0.65,0.95}{%
  \draw[accent,thick] (0.95,1.9) ++(-25:\r) arc (-25:25:\r);}

\node[person] (evehead) at (8,3.1) {};
\draw[thick] (8,3.0)--(8,2.35);
\node[align=center,font=\footnotesize] at (8,3.65) {\textbf{Eve}\\(eavesdrops)};
\draw[fill=deepgreen!25,draw=deepgreen,rounded corners=2pt] (7.35,1.5) rectangle (8.65,2.3);
\node[deepgreen,font=\scriptsize] at (8,2.05) {bag};
\draw[fill=deepgreen,draw=deepgreen] (8,1.75) circle (0.12);
\node[deepgreen,align=center,font=\scriptsize] at (8,1.15) {concealed\\microphone};

\draw[->,accent,very thick,dashed] (1.35,1.95) -- node[above,font=\scriptsize,text=accent,align=center]
  {keystroke acoustics\\(NLoS, 1--3\,m)} (7.25,1.95);

\node[draw=iiitblue,rounded corners,fill=iiitblue!8,align=center,text width=11.2cm,font=\footnotesize]
  (pipe) at (4,-0.35)
  {\textbf{Offline processing on Eve's laptop:}\ \ segmentation $\rightarrow$ \decker{} encoder
  $\rightarrow$ \topk{} keys $\rightarrow$ LLM decoding $\rightarrow$ \emph{recovered text / password}};
\draw[->,deepgreen,thick] (8,1.05) |- (pipe.east);
\end{tikzpicture}
\caption{Threat scenario. Bob types on a laptop in a public area while Eve, seated at a separate table
(1--3\,m away), passively captures keystroke acoustics from a concealed microphone with no line of sight to the
keyboard. Eve later processes the recording offline---domain-invariant acoustic inference with \decker{} followed
by LLM-driven sentence reconstruction---to recover the typed text or password.}
\label{fig:threat}
\end{figure}

We consider a common public or semi-public scenario involving a benign user, Bob, and a passive eavesdropper,
Eve (Figure~\ref{fig:threat}).

\begin{enumerate}[label=(\roman*),leftmargin=1.6em]
\item Bob uses a laptop with a built-in QWERTY keyboard while working in a library, café, classroom, or
co-working space. During typical use he types authentication credentials, emails, search queries, and
natural-language text.
\item Bob practices reasonable cybersecurity hygiene---no untrusted software, no device compromise, vigilance
against shoulder-surfing---but does not anticipate long-distance acoustic snooping using environmental
microphones.
\item Eve positions herself at a socially acceptable distance of one to three metres, often behind Bob or at a
neighbouring table. She maintains plausible behaviour and avoids close-range surveillance. Direct line of sight
to the keyboard is \emph{not} necessary.
\end{enumerate}

After recording, Eve processes the audio with the \decker{} pipeline to infer key-level sequences and then
applies a language model to reconstruct full sentences or passwords, improving inference accuracy through
linguistic context.

\section{Attacker Capabilities}
\label{sec:threat-cap}

Eve is equipped with capabilities feasible for unskilled or moderately skilled adversaries using commodity
equipment.

\paragraph{Passive acoustic capture.} Eve records ambient audio through a smartphone microphone, a compact USB
microphone, or a small microphone array. No signal transmission, active probing, or ultrasonic injection is
required.

\paragraph{No access to the victim device.} Eve does not modify Bob's software, kernel, browser, or network
stack. She cannot observe screen contents, clipboard data, or input logs. The attack is entirely external and
requires no permissions.

\paragraph{Covert placement.} The recording device is concealed inside everyday objects---a bag, a book pile, a
jacket pocket---with a footprint approximating an A4-sized object, enabling unobtrusive placement in shared
seating.

\paragraph{Non-line-of-sight robustness.} The adversary may be obstructed by Bob's hands, the laptop chassis, or
room elements. Multipath reflections off the screen, desk, and walls provide sufficient acoustic leakage for
inference (Section~\ref{sec:bg-acoustics}).

\paragraph{Offline computational resources.} Eve has a GPU-equipped or laptop-class machine for offline
execution of the domain-invariant encoder and pretrained language models (GPT-2, FLAN-T5) that correct noisy
predictions. Chapter~\ref{ch:results} shows that a commodity laptop suffices.

\paragraph{LLM-assisted post-processing.} Eve refines noisy character predictions with a language model that
exploits linguistic structure---word frequency, grammar, password patterns---elevating the attack from character
inference to full-sentence and password reconstruction.

\section{Environmental Assumptions}
\label{sec:threat-env}

The attack is assumed to occur in typical public spaces without strict acoustic control.

\begin{description}[leftmargin=1.2cm,style=nextline]
\item[Ambient noise.] Conversations, footsteps, HVAC noise, and utensil clatter are expected. Background noise
does not invalidate the attack, because keystrokes produce high-SNR transients and \decker{} employs
data-driven normalization.
\item[Multipath propagation.] Keystroke sounds reflect off the laptop screen, tabletop, and surroundings, forming
multiple arrival paths that preserve keystroke identity even when the direct path is blocked.
\item[Ordinary laptop posture.] The screen sits at typical viewing angles ($95^\circ$--$125^\circ$), creating
consistent reflection geometries.
\item[Realistic human spacing.] Eve remains at one to three metres, matching real seating patterns and avoiding
suspicious behaviour.
\end{description}

\section{Vulnerability Basis}
\label{sec:threat-vuln}

The feasibility of the attack rests on two well-established principles, stated here as properties because the
whole thesis is, in effect, an empirical test of them under realistic conditions.

\begin{property}[Distinctive, repeatable acoustic signatures]
Each keyboard key exhibits device-specific spectral and temporal patterns, and modern microphones capture enough
structured variation to discriminate among keys. This property enables key-level inference but, crucially, is
\emph{device-specific}: it is what makes single-device attacks easy and cross-device attacks hard.
\end{property}

\begin{property}[Linguistic regularity in typed text]
Even when some keystrokes are misclassified, typed sequences follow predictable linguistic constraints---bigrams,
syntax, common passwords. Language models exploit these constraints to correct local acoustic errors, enabling
recovery of semantically valid text.
\end{property}

\section{Attacker Assumptions, Summarized}
\label{sec:threat-summary}

For reference in later chapters, the threat model is captured by the following assumptions.

\begin{table}[H]
\centering
\caption{Summary of the threat model assumptions used throughout the thesis.}
\label{tab:threat-assumptions}
\small
\begin{tabularx}{\linewidth}{l X}
\toprule
\textbf{ID} & \textbf{Assumption} \\
\midrule
A1 & Passive recording only; no probes or malware. \\
A2 & Commodity hardware microphones. \\
A3 & Concealed placement within everyday objects. \\
A4 & Socially realistic distance (1--3\,m). \\
A5 & NLoS operation enabled by multipath reflections. \\
A6 & Robustness to ambient real-world noise. \\
A7 & Offline laptop- or GPU-class computation permitted. \\
A8 & Access to pretrained LLMs for decoding sequences. \\
\bottomrule
\end{tabularx}
\end{table}

This threat model captures both the physical and algorithmic realism of modern \asca{} attacks, including the
enhanced inference potential unlocked by LLM-guided reconstruction. Every accuracy number reported in this
thesis should be read against it: a result is only meaningful to the extent that it holds when Eve does not
control Bob's hardware, cannot fix the recording geometry, and must contend with noise, multipath, and unseen
keyboards.

\chapter{The HEAR Dataset}
\label{ch:hear}

Robust evaluation of cross-device \asca{} requires a dataset that reflects the diversity of real-world users,
keyboards, microphone types, and environments. Prior datasets have been limited in scope, typically consisting
of single-keyboard, single-user, clean-environment collections, which makes it difficult to study whether \asca{}
can generalize beyond laboratory conditions. To address this, we introduce \hear{}, a multi-device, multi-user,
multi-gender, multi-environment corpus with synchronized recordings across laptops, smartphones, and streamed
audio channels (VoIP). The dataset is collected using a reproducible browser-based acquisition pipeline and is the
most comprehensive \asca{} resource to date. \hear{} is the single dataset used throughout this thesis---for the
cross-keyboard experiments of Chapter~\ref{ch:results} and the VoIP experiments of Chapter~\ref{ch:keyac} alike---and
was released with the \decker{} framework~\cite{maurya2026decker}. It is publicly documented at
\url{https://github.com/vikrant-vikram/Hear-Dataset}.

\section{Design Goals}
\label{sec:hear-goals}

\hear{} is constructed around six core objectives, each of which is essential for realistic \asca{} research and
each of which maps to a specific evaluation axis in Chapter~\ref{ch:results}.

\begin{enumerate}[label=(\roman*),leftmargin=1.7em]
\item \textbf{Keyboard diversity:} laptop scissor-switch keyboards, external membrane keyboards, and various
mechanical-switch boards, so that device impulse responses span a wide range.
\item \textbf{User diversity:} participants across genders, handedness, and typing backgrounds, with metadata
captured in-session, enabling analysis of user bias.
\item \textbf{Environmental realism:} recordings in quiet (library), moderate-noise (office), and high-noise
(café/outdoor) settings.
\item \textbf{Device diversity:} built-in laptop microphones, smartphones placed at varying distances and
orientations, and VoIP/streamed audio.
\item \textbf{Synchronization for multi-recorder \asca{}:} explicit offset estimation between primary and
secondary devices, enabling multi-channel analysis.
\item \textbf{Reproducibility:} every component is automated and recorded with structured metadata.
\end{enumerate}

This design enables controlled cross-device, cross-user, and cross-environment splits, which are essential for
evaluating domain-invariant \asca{} models such as \decker{}.

\section{Comparison with Prior ASCA Datasets}
\label{sec:hear-compare}

\hear{} is the first dataset to unify all of the \asca{}-relevant factors---particularly multi-gender,
over-stream, multiple-microphone, NLoS, and continuous-text typing---yielding the broadest known \asca{}
evaluation surface. This diversity is crucial for evaluating generalization-driven methods such as \decker{}.
Table~\ref{tab:dataset-compare} in Chapter~\ref{ch:related} sets \hear{} against prior datasets and makes the
gap concrete: earlier collections vary one or two factors, whereas \hear{} varies seven jointly, at a scale of
more than ten thousand keystroke events.

\section{Acquisition Pipeline}
\label{sec:hear-pipeline}

All data were collected using a custom browser-based tool written in HTML and JavaScript and provided as part of
the research artifact. The tool is designed to (i) capture continuous audio using the \texttt{MediaRecorder} API,
(ii) log high-resolution keystroke timestamps, (iii) record participant metadata (gender, handedness, session
number, and so on), and (iv) export one continuous audio file plus structured metadata and pre-sliced per-key WAV
clips. The interface also presents an informed-consent checkbox and captures the demographic fields essential for
user-diversity analysis. Because acquisition runs in a browser, it is portable across laptops without installing
software, which is what makes a thirty-seven-keyboard collection feasible.

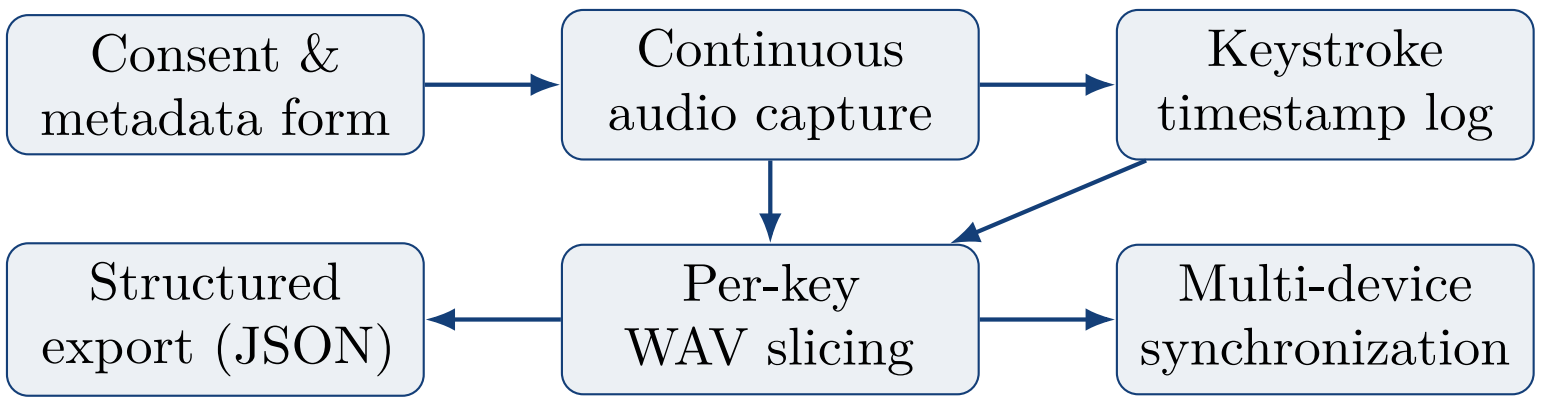
\begin{figure}[t]
\centering
\begin{tikzpicture}[font=\small,node distance=0.55cm and 0.9cm,
  box/.style={draw=iiitblue,fill=iiitblue!8,rounded corners,minimum height=0.9cm,text width=2.5cm,align=center},
  arr/.style={-{Latex[length=2mm]},thick,iiitblue}]
\node[box] (consent) {Consent \&\\metadata form};
\node[box,right=of consent] (rec) {Continuous\\audio capture};
\node[box,right=of rec] (ts) {Keystroke\\timestamp log};
\node[box,below=of rec] (slice) {Per-key\\WAV slicing};
\node[box,right=of slice] (sync) {Multi-device\\synchronization};
\node[box,left=of slice] (export) {Structured\\export (JSON)};
\draw[arr] (consent)--(rec);
\draw[arr] (rec)--(ts);
\draw[arr] (rec)--(slice);
\draw[arr] (ts)--(slice);
\draw[arr] (slice)--(sync);
\draw[arr] (slice)--(export);
\end{tikzpicture}
\caption{The \hear{} browser-based acquisition pipeline. A single session yields a continuous recording, a
high-resolution keystroke timestamp log, pre-sliced per-key clips, structured metadata, and, where secondary
devices are present, synchronization offsets.}
\label{fig:pipeline}
\end{figure}

\section{Standardized Typing Corpus}
\label{sec:hear-corpus}

Participants are provided with a standardized paragraph crafted to cover the alphabet (A--Z, a--z), digit
patterns, punctuation, symbols and operators, uppercase acronyms, and realistic word and phrase frequencies for
language-model reconstruction. Participants also typed a structured block of special keys (Tab, Backspace, Delete,
arrows, Esc, F1--F12, Caps Lock, Enter), enabling non-character key classification experiments. The full corpus is
reproduced in Appendix~\ref{app:corpus}. Standardizing the corpus serves two purposes: it guarantees that every
key is exercised enough times to train and evaluate a classifier, and it fixes the linguistic distribution so that
language-model decoding is evaluated against a known and realistic text rather than an artificially easy or hard one.

\section{Per-Key Slicing and Signal Processing}
\label{sec:hear-slicing}

Given a keystroke timestamp $t_{\text{ms}}$, audio is sliced using the fixed asymmetric window of
Equation~\eqref{eq:window}, reproduced here for convenience:
\begin{equation}
t_{\text{start}} = t_{\text{ms}} - 60\,\text{ms}, \qquad t_{\text{end}} = t_{\text{ms}} + 200\,\text{ms}.
\end{equation}
This window captures both the impulsive onset and early reverberation of keystrokes, consistent with prior \asca{}
literature. Windows shorter than a fixed minimum are discarded to remove empty or noise-only segments. Rapid typing
can cause partial overlap between adjacent windows; such overlap is retained intentionally to reflect realistic
attack conditions, where a passive attacker cannot cleanly separate fast keystrokes.

\section{Multi-Device and Over-Stream Synchronization}
\label{sec:hear-sync}

Because \hear{} includes smartphone audio, VoIP feeds, and secondary microphones, synchronization is performed
relative to the browser reference recording. When explicit synchronization tones are available, they allow
estimation of an offset $\hat{o}_d$ and a linear drift $b_d$ for device $d$:
\begin{equation}
t_d(t) = t + \hat{o}_d + b_d\, t.
\end{equation}
When markers are unavailable, the offset is estimated via normalized cross-correlation between the reference signal
$x$ and the device signal $y$:
\begin{equation}
\hat{\tau} = \arg\max_{\tau} \sum_t x(t)\,y(t+\tau),
\end{equation}
and drift is estimated using correlations over temporally separated segments. Residual alignment errors are typically
within a few milliseconds, which we account for in the segmentation-robustness experiments of Chapter~\ref{ch:results}.
Accurate synchronization is what turns \hear{} from a collection of independent recordings into a genuinely
multi-channel resource, in which the \emph{same} keystroke is observed simultaneously through several very different
acoustic paths---exactly the setting that exposes, and then lets us suppress, device colouration.

\section{Metadata Schema}
\label{sec:hear-metadata}

We collect metadata from users in each session, comprising the continuous browser recording (WebM/OGG), per-key
WAVs for all devices, and complete metadata in JSON. The schema preserves demographic, acoustic, environmental, and
segmentation-relevant information required by downstream modeling; a representative record is given in
Appendix~\ref{app:schema}. Storing demographics alongside the audio is what enables the user-bias analysis that no
prior \asca{} dataset supports: the influence of gender, age group, and hand dominance on typing speed, key pressure,
and error rate, and hence on attack success.

\section{Dataset Statistics}
\label{sec:hear-stats}

\hear{} comprises more than ten thousand labelled keystroke events drawn from $53$ participants and $37$ laptop
keyboards, recorded across three channels and three environment classes. Table~\ref{tab:hear-stats} summarizes the
headline counts along the axes that most affect generalization: gender, recording channel, and environmental noise.
Two design choices are visible in the numbers.
First, the near-balanced gender split is deliberate and, to our knowledge, unique among \asca{} datasets; it is what
makes the user-bias analysis of Chapter~\ref{ch:discussion} possible. Second, the three recording channels are
balanced by construction, so that no single acoustic path dominates training and the cross-device split is not
confounded by sample-count imbalance.

\begin{table}[h]
\centering
\caption{Headline statistics of the \hear{} dataset.}
\label{tab:hear-stats}
\small
\begin{tabularx}{\linewidth}{l X}
\toprule
\textbf{Attribute} & \textbf{Value} \\
\midrule
Participants & $53$ (near gender-balanced) \\
Laptop keyboards & $37$ (scissor, membrane, mechanical) \\
Recording channels & external microphone, on-device microphone, VoIP stream \\
Environments & quiet (library), moderate (office), high-noise (café/outdoor) \\
Labelled keystroke events & $>10{,}000$ per primary channel \\
Character classes & $94$ printable + special keys (Tab, Enter, arrows, F1--F12, \dots) \\
Sampling rate & $48$\,kHz (resampled to $16$\,kHz for PTM features) \\
Per-key window & $[-60,\,+200]$\,ms around onset \\
Metadata & gender, handedness, age group, typing background, awareness \\
\bottomrule
\end{tabularx}
\end{table}

\section{Recommended Evaluation Splits}
\label{sec:hear-splits}

We define four canonical splits for reproducibility, each isolating one axis of generalization.

\begin{description}[leftmargin=1.2cm,style=nextline]
\item[Cross-Keyboard (primary).] Train on a subset of keyboards; test on unseen keyboards. This is the primary
axis, because an attacker rarely controls the victim's hardware.
\item[Cross-User.] Disjoint user identities between train and test, isolating typist-specific variation.
\item[Cross-Environment.] Train on quiet/moderate settings, test on café/outdoor, isolating ambient-noise and
room effects.
\item[Cross-Device.] Train on laptop audio, test on smartphone/VoIP, isolating microphone and channel effects.
\end{description}

Metrics include key-level accuracy, \topk{} accuracy, sentence-level accuracy, Levenshtein edit distance, and
segmentation robustness, defined precisely in Chapter~\ref{ch:results}. The splits are deliberately \emph{leave-one-
domain-out} rather than random, because random splits leak device identity across train and test and produce the
inflated numbers that this thesis argues are misleading.

\section{Ethics, Consent, and User Awareness}
\label{sec:hear-ethics}

We follow established ethical procedures throughout. All participants provided informed consent before
participating. Personally identifiable information is removed with explicit care, and we consider $k$-anonymity on
the data before any experimental use. Data access is limited to requests for academic research purposes only. In
addition to the acoustic recordings, we captured participants' \emph{awareness} of acoustic leakage threats, which
provides a small but novel window into the human factor: most participants reported no awareness that their typing
could be recovered acoustically, which is precisely the condition under which a passive attack succeeds. The
combination of realistic data, careful consent, and awareness annotations is what allows \hear{} to support an honest
study of \asca{} generalization across gender, device type, microphone type, VoIP channel, and environmental
noise---an essential foundation for evaluating modern \asca{} pipelines such as \decker{}.

\keybox{Chapter takeaway.}{\hear{} supplies the missing ingredient in \asca{} research: realistic, diverse,
synchronized, ethically collected data whose leave-one-domain-out splits make cross-keyboard generalization
measurable rather than assumed.}

\chapter{Acoustic Side-Channel Attacks over VoIP}
\label{ch:keyac}

The \decker{} framework of this thesis (Chapter~\ref{ch:decker}) targets the primary open-world axis:
generalization across unseen keyboards. A second axis, equally important for realistic attacks, is the
\emph{transmission channel}. A large and growing fraction of sensitive typing now happens during online
meetings, where the audio is compressed, gain-controlled, and echo-cancelled by a Voice-over-IP (VoIP) pipeline
before an eavesdropper on the call ever hears it. This chapter studies whether keystroke inference survives that
channel, using the VoIP recordings contained in the \hear{} dataset (Chapter~\ref{ch:hear}). It proceeds in two
complementary parts. Part~one asks how well modern \emph{speech representations} transfer to codec-distorted
keystrokes, and shows that an expressive Kolmogorov--Arnold adaptation head recovers much of the accuracy lost to
compression. Part~two goes further and reformulates the attack itself as \emph{continuous sparse transduction} of
an unsegmented stream, introducing \evflow{}, which canonicalizes distorted VoIP embeddings before decoding. Both
parts are evaluated on the same \hear{} data, and both are framed as robustness tools that complement---rather than
replace---the domain-invariant \decker{} model that remains the centrepiece of the thesis.

\section{Motivation}
\label{sec:keyac-motiv}

VoIP is not a corner case. Codecs apply automatic gain control (AGC) and acoustic echo cancellation (AEC), and they
compress spectrally and temporally in ways tuned for speech intelligibility, not for the impulsive, broadband
transients of a keystroke. Two questions follow. First, do the speech pretrained models that increasingly serve as
\asca{} front-ends still provide usable representations once the signal has passed through a codec? Second, is the
standard segment-then-classify pipeline itself the right abstraction for a continuous, passively captured VoIP
stream, in which keystrokes are sparse events separated by long stretches of silence and background noise? This
chapter answers both, entirely on the VoIP portion of \hear{}, so that no additional dataset is introduced.

\section{The HEAR VoIP Subset}
\label{sec:keyac-data}

All experiments in this chapter use the VoIP recordings of \hear{}. As described in Chapter~\ref{ch:hear}, each
keystroke in \hear{} is captured simultaneously through a local laptop microphone, a mobile device, and a real-time
VoIP pipeline (Zoom, Microsoft Teams, and Google Meet). The VoIP evaluation subset comprises $12{,}480$ synchronized
micro-acoustic events drawn from the same $53$ participants and $37$ laptop keyboards as the rest of \hear{}, with
ground-truth labels and temporal boundaries obtained from the synchronized keystroke log. Because the local-microphone
and VoIP recordings of each keystroke are \emph{frame-aligned} (Chapter~\ref{ch:hear}, Section~\ref{sec:hear-sync}),
\hear{} provides paired clean/distorted views of the same event---a property that Part~two exploits directly. To
prevent leakage we use a strict disjoint train/validation/test split ($70\%/10\%/20\%$) in which the test set
contains entirely unseen users and keyboards, so the evaluation measures hardware and typing-style generalization
rather than user-specific memorization. The dataset and its VoIP subset are publicly documented at
\url{https://github.com/vikrant-vikram/Hear-Dataset}.

\section{Part I: Speech Representations under VoIP}
\label{sec:keyac-ptm}

\subsection{Pretrained Models Evaluated}
We evaluate six representations, introduced in Section~\ref{sec:bg-ptm}. Four are self-supervised: Wav2Vec2, WavLM,
and HuBERT ($768$-dimensional), and XLS-R ($1024$-dimensional). Two are supervised: X-vectors and the Whisper encoder
($512$-dimensional); for Whisper we use only the encoder to extract embeddings. All backbones are frozen throughout,
and audio is resampled to $16$\,kHz before feature extraction. Freezing isolates the intrinsic quality of the
representation for the impulsive-sound domain, which is far from the clean-speech domain the backbones were trained on.

\subsection{Adaptation Heads}
We compare three heads over the frozen embeddings. A \textbf{fully connected network (FCN)} with a $90$-neuron hidden
layer is the minimal baseline. A \textbf{convolutional network (CNN)} with two convolution/max-pool blocks captures
local structure. The proposed \textbf{Kolmogorov--Arnold Network (KAN)} head (Section~\ref{sec:bg-kan}) uses a single
hidden KAN layer with $30$ units, grid size $5$, and spline order $3$, replacing linear transformations with learnable
univariate spline functions so that the head models nonlinear feature interactions explicitly. The hypothesis is that
the representation drift induced by codec compression is strongly nonlinear and therefore poorly captured by linear or
shallow heads. All heads are trained with cross-entropy, AdamW, $20$ epochs, batch size $16$, learning rate
$2.2\times10^{-4}$, with dropout and early stopping. This frozen-backbone, lightweight-head protocol follows
established downstream-modeling practice in speech and multimodal representation
learning~\cite{phukan2025rethinking,choudhury2026foca,choudhury2026gocoma}.

\subsection{Protocol and Results}
We evaluate under in-domain (ID, $5$-fold cross-validation) and out-of-domain (OOD) settings, across three scenarios
that disentangle keyboard variation from codec distortion: keyboard generalization on standard recordings, keyboard
generalization on VoIP-streamed audio of held-out keyboards, and codec generalization (train on standard, test on
VoIP). Table~\ref{tab:keyac} reports the full grid; the WavLM row is the strongest throughout.

\paragraph{Backbone ranking.} Among the pretrained models, WavLM consistently achieves the highest performance. With
the CNN head it reaches $58.34\%$ accuracy and $56.92\%$ mF1 in the in-domain standard setting, outperforming the other
backbones by a clear margin, while HuBERT and XLS-R trail at $36.27\%$ and $33.45\%$. WavLM's denoising objective, which
exposes it to masked and corrupted speech during pretraining, appears to yield representations better matched to the
noisy, transient character of keystroke sounds. Under keyboard holdout, all models lose roughly $8$--$10$ points.

\paragraph{The VoIP penalty and the KAN advantage.} Under VoIP conditions performance falls further due to
codec-induced distortion, but the proposed KAN head improves performance across every backbone and scenario. With
WavLM it reaches $68.47\%$ accuracy in-domain and \emph{maintains} $61.73\%$ under keyboard holdout, and the OOD
degradation under codec shift is markedly smaller than for the baselines. The representations that a linear head throws
away under codec distortion are recoverable by a head expressive enough to model the nonlinear warp. This study is
reported in the companion publication on speech-representation impact analysis~\cite{choudhury2026keyac}.

\begin{table}[t]
\centering
\caption{Generalization performance on the \hear{} VoIP subset, reported as Accuracy(\%) / Macro-F1(\%). Best result in
each column is in \textbf{bold}. The proposed KAN head dominates across all backbones and scenarios.}
\label{tab:keyac}
\scriptsize
\setlength{\tabcolsep}{3.2pt}
\begin{tabular}{l cc cc cc cc cc cc}
\toprule
& \multicolumn{4}{c}{\textbf{Keyboard Gen.\ (Standard)}} & \multicolumn{4}{c}{\textbf{Keyboard Gen.\ (VoIP)}} & \multicolumn{4}{c}{\textbf{Codec Gen.}}\\
\cmidrule(lr){2-5}\cmidrule(lr){6-9}\cmidrule(lr){10-13}
& \multicolumn{2}{c}{ID} & \multicolumn{2}{c}{OOD} & \multicolumn{2}{c}{ID} & \multicolumn{2}{c}{OOD} & \multicolumn{2}{c}{ID} & \multicolumn{2}{c}{OOD}\\
Model & Acc & mF1 & Acc & mF1 & Acc & mF1 & Acc & mF1 & Acc & mF1 & Acc & mF1 \\
\midrule
\multicolumn{13}{l}{\textit{CNN Downstream}}\\
Wav2Vec2 & 47.82 & 46.41 & 38.95 & 37.64 & 33.71 & 32.26 & 25.14 & 23.88 & 34.15 & 32.78 & 26.54 & 25.18 \\
WavLM    & 58.34 & 56.92 & 49.47 & 48.16 & 44.12 & 42.73 & 35.86 & 34.54 & 45.21 & 43.86 & 37.54 & 36.21 \\
HuBERT   & 36.27 & 34.84 & 27.61 & 26.18 & 25.92 & 24.71 & 18.66 & 17.42 & 26.83 & 25.52 & 19.44 & 18.21 \\
XLS-R    & 33.45 & 32.08 & 24.82 & 23.47 & 23.41 & 22.24 & 16.75 & 15.58 & 24.12 & 22.91 & 17.36 & 16.04 \\
X-Vectors& 44.18 & 42.86 & 35.74 & 34.21 & 30.27 & 28.91 & 22.65 & 21.36 & 31.28 & 29.91 & 23.76 & 22.41 \\
Whisper  & 48.73 & 47.41 & 39.96 & 38.54 & 35.64 & 34.31 & 27.54 & 26.21 & 36.47 & 35.14 & 28.62 & 27.23 \\
\midrule
\multicolumn{13}{l}{\textit{FCN Downstream}}\\
Wav2Vec2 & 45.36 & 43.92 & 36.54 & 35.11 & 31.68 & 30.24 & 23.94 & 22.71 & 32.47 & 31.08 & 24.82 & 23.54 \\
WavLM    & 55.81 & 54.36 & 46.94 & 45.52 & 40.35 & 39.02 & 32.48 & 31.14 & 41.72 & 40.41 & 33.86 & 32.44 \\
HuBERT   & 34.41 & 33.08 & 26.02 & 24.78 & 23.85 & 22.64 & 17.21 & 16.02 & 24.73 & 23.51 & 17.92 & 16.67 \\
XLS-R    & 31.74 & 30.36 & 23.42 & 22.11 & 21.64 & 20.43 & 15.02 & 13.88 & 22.56 & 21.27 & 15.94 & 14.72 \\
X-Vectors& 41.63 & 40.18 & 33.21 & 31.82 & 28.26 & 26.91 & 20.73 & 19.41 & 29.54 & 28.16 & 22.11 & 20.74 \\
Whisper  & 46.25 & 44.91 & 37.58 & 36.17 & 33.14 & 31.86 & 25.27 & 23.96 & 34.08 & 32.76 & 26.42 & 25.07 \\
\midrule
\multicolumn{13}{l}{\textit{KAN Downstream (Proposed)}}\\
Wav2Vec2 & 66.18 & 64.82 & 59.74 & 58.33 & 55.91 & 54.62 & 49.36 & 48.02 & 56.72 & 55.41 & 51.28 & 49.93 \\
\rowcolor{softgray}WavLM & \textbf{68.47} & \textbf{67.12} & \textbf{61.73} & \textbf{60.34} & \textbf{57.82} & \textbf{56.41} & \textbf{50.91} & \textbf{49.56} & \textbf{58.36} & \textbf{57.04} & \textbf{52.84} & \textbf{51.46} \\
HuBERT   & 63.28 & 61.94 & 56.47 & 55.11 & 52.06 & 50.73 & 46.23 & 44.91 & 52.18 & 50.86 & 47.21 & 45.83 \\
XLS-R    & 61.74 & 60.42 & 54.93 & 53.62 & 50.47 & 49.12 & 44.51 & 43.22 & 50.96 & 49.64 & 46.08 & 44.71 \\
X-Vectors& 58.41 & 57.06 & 51.62 & 50.23 & 47.18 & 45.86 & 41.63 & 40.32 & 48.17 & 46.83 & 42.96 & 41.52 \\
Whisper  & 65.74 & 64.39 & 59.12 & 57.83 & 54.83 & 53.46 & 48.92 & 47.63 & 55.63 & 54.28 & 50.41 & 49.06 \\
\bottomrule
\end{tabular}
\end{table}

\section{Part II: Continuous Sparse Transduction with EvFlow-CTSSM}
\label{sec:keyac-evflow}

Part~one keeps the conventional segment-then-classify pipeline. But a fully passive VoIP attacker records an
\emph{unsegmented} stream in which keystrokes are a sparse temporal point process---a handful of impulsive events per
second amid long stretches of silence---and codecs distort that stream non-linearly. Two problems follow: fixed-window
classifiers are brittle to boundary jitter, and standard self-attention wastes $O(T^2)$ capacity attending to
irrelevant silence. We therefore reformulate the attack as \emph{continuous sparse transduction} and introduce
\evflow{} (Event-Triggered Continuous-Time State-Space Modeling with Latent Flow Canonicalization), evaluated on the
same \hear{} VoIP subset~\cite{maurya2026evflow}.

\subsection{Problem Formulation}
Let $X_{\text{voip}}\in\real^{T\times F}$ be the unsegmented, distorted acoustic feature sequence (log-mel filterbanks)
of length $T$, and $Y=(y_1,\dots,y_U)$ the target event sequence of length $U$, with $U\ll T$. \evflow{} decodes this
sparse point process in three phases: latent flow canonicalization, event-triggered state modeling, and CTC decoding.

\subsection{Latent Rectified Flow Canonicalization (LF-CAN)}
Because VoIP pipelines apply dynamic, non-linear distortion (AEC and AGC), deterministic waveform inversion is
ill-posed. \lfcan{} instead performs zero-shot canonicalization in the \emph{latent} space using rectified flow
matching~\cite{liu2025flowmatching}. The frame-aligned clean/VoIP pairs of \hear{} yield paired latent embeddings $Z_{\text{voip}}, Z_{\text{clean}}\in\real^{T\times D}$
from a frozen shallow extractor; this weak supervision learns the manifold mapping and is never required at inference.
A linear probability path is constructed between the distorted distribution ($Z_0=Z_{\text{voip}}$) and the clean
canonical distribution ($Z_1=Z_{\text{clean}}$), and a neural vector field $v_\theta(Z_t,t)$ parameterizes an ODE
\begin{equation}
dZ_t = v_\theta(Z_t,t)\,dt,
\end{equation}
trained to match the linear interpolation between the paired distributions by minimizing the transport cost
\begin{equation}
\Loss_{\text{flow}} = \mathbb{E}_{t\sim\mathcal{U}[0,1],\,Z_0,Z_1}\Big[\big\|v_\theta(Z_t,t)-(Z_1-Z_0)\big\|_2^2\Big].
\end{equation}
At inference, \lfcan{} solves the ODE forward in a single Euler step to yield canonicalized latents $Z_{\text{clean}}$.
The effect is analogous to the device-coloration suppression that KSN achieves in \decker{}
(Chapter~\ref{ch:results}, Section~\ref{sec:results-ksn}): channel-induced drift is collapsed onto a common manifold.

\subsection{Event-Triggered Continuous-Time SSM}
Standard attention scales as $O(T^2)$ and spends uniform capacity across time, which is wasteful for sparse impulses.
We replace it with a data-dependent continuous-time state-space model, in the spirit of selective state spaces~\cite{gu2024mamba}.
The base SSM maps input to a hidden state,
$\dot{h}(t)=A h(t)+B Z_{\text{clean}}(t)$, $y(t)=C h(t)$. To allocate capacity strictly to acoustic impulses we
introduce a continuous event-trigger gate $g(t)=\sigma(W_g Z_{\text{clean}}(t)+b_g)\in[0,1]$ that modulates the
discretized integration step,
\begin{equation}
\bar{A}_t = \exp\!\big(g(t)\,\Delta_t\,A\big), \qquad \bar{B}_t = (\bar{A}_t - I)A^{-1}B, \qquad h_t = \bar{A}_t h_{t-1} + \bar{B}_t Z_{\text{clean}}(t).
\end{equation}
When $g(t)\approx 0$ (silence) the transition approaches the identity and the hidden state is preserved without
integrating ambient noise; when $g(t)\approx 1$ (a transient) the model rapidly updates its state, localizing
integration to acoustic events.

\subsection{Decoding via CTC and Shallow Fusion}
The CT-SSM output is projected to the vocabulary (plus a blank) and trained end-to-end with the Connectionist Temporal
Classification (CTC) loss~\cite{graves2006ctc}, $\Loss_{\text{CTC}} = -\log\sum_{A\in\mathcal{B}^{-1}(Y)}\prod_{t=1}^{T} p(a_t\mid X_{\text{voip}})$,
where $\mathcal{B}$ is the CTC collapse function. To separate acoustic transduction from linguistic memorization we use
formal shallow fusion during beam search, $Y^{*}=\arg\max_Y\big(\log P_{\text{CTC}}(Y\mid X_{\text{voip}})+\lambda\log P_{\text{LM}}(Y)\big)$,
and by controlling $\lambda$ we isolate the true acoustic resolving power of the model from the language prior---the same
acoustic-versus-linguistic separation that motivates the LLM analysis of \decker{} (Chapter~\ref{ch:results}).

\subsection{Results}
Table~\ref{tab:evflow} reports continuous character error rate (CER) on unseen keyboards and VoIP channels. SSL
foundation models and transformer AED models achieve low CER on seen domains but suffer severe degradation ($+28$ to
$+33\%$ CER) on unseen VoIP pipelines---including the Audio Spectrogram Transformer~\cite{gong2021ast}---and standard
ASR Conformers fail on mechanical transients. \evflow{} degrades by only $+11.1\%$, demonstrating robust physical
decoding under non-linear AGC suppression. An information-theoretic analysis further lower-bounds the mutual
information recovered per keystroke, indicating that the gain reflects genuine acoustic transduction rather than
language-model memorization~\cite{cover2006infotheory}.

\begin{table}[h]
\centering
\caption{Continuous CER ($\downarrow$) on unseen keyboards and VoIP channels (\hear{} VoIP subset). \evflow{} is compared
against ASR, AED, and SSL foundation baselines.}
\label{tab:evflow}
\small
\begin{tabular}{lccc}
\toprule
Model architecture & Seen-domain CER & Unseen-domain CER & Degradation \\
\midrule
\multicolumn{4}{l}{\textit{Standard ASR \& fixed-window AED}}\\
Conformer-CTC & 22.4\% & 54.1\% & $+31.7$ \\
CRNN (AED) & 17.8\% & 48.9\% & $+31.1$ \\
ECAPA-TDNN & 18.2\% & 47.6\% & $+29.4$ \\
\multicolumn{4}{l}{\textit{SSL audio foundation models}}\\
wav2vec 2.0 & 14.8\% & 48.2\% & $+33.4$ \\
HuBERT & 13.5\% & 46.5\% & $+33.0$ \\
WavLM & 12.9\% & 44.1\% & $+31.2$ \\
\multicolumn{4}{l}{\textit{Transformer AED}}\\
AST & 15.1\% & 43.8\% & $+28.7$ \\
\rowcolor{softgray}\evflow{} (ours) & 16.4\% & \textbf{27.5\%} & $\mathbf{+11.1}$ \\
\bottomrule
\end{tabular}
\end{table}

\paragraph{Acoustic-only (gibberish) benchmark.} To decouple acoustic resolving power from linguistic memorization we
evaluate CER on high-entropy random strings while modulating the shallow-fusion weight $\lambda$. At $\lambda=0$ (pure
acoustic transduction) \evflow{} yields $31.2\%$ CER on gibberish, showing it physically resolves impulses without
linguistic priors; as $\lambda$ rises to $0.5$, structured-text CER drops to $8.1\%$ while gibberish CER remains near
$27.9\%$, confirming the CT-SSM acts as a robust physical anchor rather than a language-model memorizer. Under
$\pm30$--$50$\,ms boundary jitter on unsegmented streams, the continuous-time model remains stable where fixed-window
baselines degrade sharply, and across disjoint splits its CER rises only modestly from $13.1\%$ (seen user/keyboard) to
$16.4\%$ (unseen user/keyboard).

\section{Relation to DECKER}
\label{sec:keyac-relation}

The two parts of this chapter and the \decker{} framework attack the same underlying problem---representation drift---from
three angles. \decker{} (Chapter~\ref{ch:decker}) removes \emph{device} coloration to generalize across keyboards; the
KAN head of Part~one recovers accuracy lost to \emph{codec} distortion with a more expressive adaptation layer; and
\evflow{} of Part~two canonicalizes \emph{channel} distortion in the latent space and decodes the resulting sparse stream
continuously. All three are evaluated on \hear{}, and a complete attacker would combine them. \decker{} remains the
central contribution of the thesis; the VoIP results here show that its domain-invariance principle extends naturally to
the transmission channel.

\keybox{Chapter takeaway.}{On \hear{}'s VoIP subset, WavLM is the strongest representation, a KAN adaptation head recovers
much of the codec penalty, and \evflow{} makes fully continuous, segmentation-free VoIP attacks viable---each reinforcing
\decker{}'s core lesson that suppressing domain and channel drift is what makes \asca{} generalize.}

\chapter{DECKER: A Domain-Invariant Inference Framework}
\label{ch:decker}

This chapter presents \decker{}---\emph{Domain-invariant Embedding for Cross-Keyboard Extraction and
Recognition}~\cite{maurya2026decker}---in full detail. \decker{} is the central contribution of this thesis, and
the framework and the accompanying \hear{} dataset were published at ACM ASIACCS~2026~\cite{maurya2026decker}. \decker{} is designed to meet two requirements that Chapters~\ref{ch:intro}
and~\ref{ch:background} identified as necessary for real-world \asca{}: at the \emph{acoustic level}, it must learn
representations that are discriminative for key identity while invariant to keyboard, user, and environment; and at the
\emph{sequence level}, it must use a language model to correct local key errors and reconstruct meaningful text or
passwords. We first give a mathematical formulation of the problem, then describe each component of the framework in
turn, and finally specify the end-to-end training and inference pipeline.

\section{Problem Definition}
\label{sec:decker-problem}

We consider keystroke samples $\Dset = \{(x_i, y_i, d_i)\}_{i=1}^{N}$, where $x_i \in \real^{T}$ is the raw
waveform, $y_i \in \{1,\dots,|K|\}$ is the key label, and $d_i \in \{1,\dots,|\Dset|\}$ denotes the keyboard domain
identity. Let $\Enc(\cdot)$ be the encoder producing embeddings $z_i = \Enc(x_i) \in \real^{E}$. \decker{} aims to
learn an embedding space satisfying three properties.

\begin{definition}[Key discriminability]
The embedding must retain enough information to predict $y_i$: there exists a classifier $C_k$ with high accuracy on
$P(y\mid z)$.
\end{definition}

\begin{definition}[Keyboard invariance]
The embedding must be uninformative about $d_i$: the best possible domain classifier $C_d$ on $P(d\mid z)$ should be
near chance.
\end{definition}

\begin{definition}[Cross-keyboard consistency]
Embeddings of the same key on \emph{different} keyboards must be close: for $y_i = y_j$ with $d_i \neq d_j$, the
distance $\|z_i - z_j\|$ should be small relative to embeddings of different keys.
\end{definition}

Together these three properties are what enable generalization to unseen devices. Note that they are in tension:
maximizing key discriminability alone recovers the overfit baselines of Chapter~\ref{ch:intro}, because on any fixed
training set the keyboard identity is a \emph{useful} feature for predicting the key. Invariance and consistency are
the constraints that forbid the model from taking that shortcut.

\section{Overview of the Pipeline}
\label{sec:decker-overview}

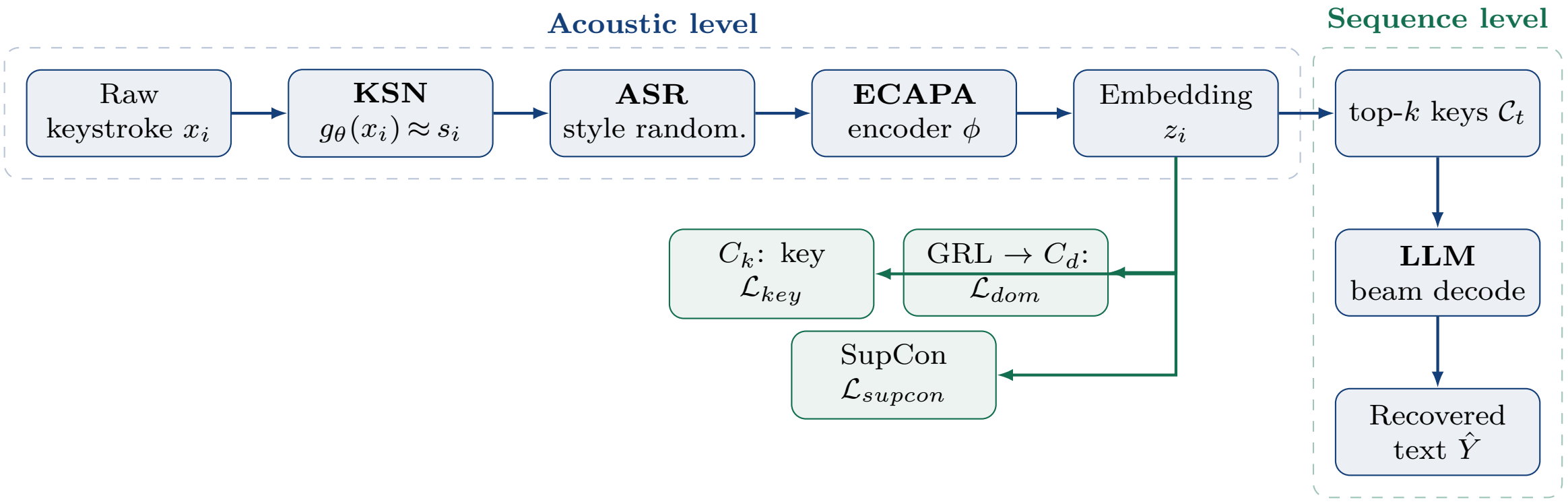
\begin{figure}[t]
\centering
\begin{tikzpicture}[font=\scriptsize,node distance=0.42cm and 0.55cm,
  b/.style={draw=iiitblue,fill=iiitblue!8,rounded corners,minimum height=0.85cm,text width=1.75cm,align=center},
  t/.style={draw=deepgreen,fill=deepgreen!8,rounded corners,minimum height=0.85cm,text width=1.75cm,align=center},
  a/.style={-{Latex[length=1.6mm]},thick,iiitblue}]
\node[b] (raw) {Raw keystroke $x_i$};
\node[b,right=of raw] (ksn) {\textbf{KSN}\\ $g_\theta(x_i)\!\approx\! s_i$};
\node[b,right=of ksn] (asr) {\textbf{ASR}\\ style random.};
\node[b,right=of asr] (enc) {\textbf{ECAPA}\\ encoder $\Enc$};
\node[b,right=of enc] (emb) {Embedding $z_i$};
\draw[a] (raw)--(ksn); \draw[a] (ksn)--(asr); \draw[a] (asr)--(enc); \draw[a] (enc)--(emb);
\node[t,below=0.7cm of enc,xshift=-1.4cm] (key) {$C_k$: key $\Loss_{key}$};
\node[t,below=0.7cm of enc,xshift=0.9cm] (dom) {GRL $\to C_d$: $\Loss_{dom}$};
\node[t,below=1.7cm of enc,xshift=-0.2cm] (con) {SupCon $\Loss_{supcon}$};
\draw[a,deepgreen] (emb) |- (key); \draw[a,deepgreen] (emb) |- (dom); \draw[a,deepgreen] (emb) |- (con);
\node[b,right=of emb] (topk) {\topk{} keys $\mathcal{C}_t$};
\node[b,below=0.7cm of topk] (llm) {\textbf{LLM}\\ beam decode};
\node[b,below=0.7cm of llm] (out) {Recovered text $\hat{Y}$};
\draw[a] (emb)--(topk); \draw[a] (topk)--(llm); \draw[a] (llm)--(out);
\begin{scope}[on background layer]
\node[draw=iiitblue!30,dashed,rounded corners,fit=(raw)(emb),inner sep=6pt,label={[iiitblue]above:\textbf{Acoustic level}}]{};
\node[draw=deepgreen!40,dashed,rounded corners,fit=(topk)(out),inner sep=6pt,label={[deepgreen]above:\textbf{Sequence level}}]{};
\end{scope}
\end{tikzpicture}
\caption{The \decker{} pipeline. Raw keystrokes are normalized (KSN), augmented (ASR), and encoded (ECAPA-TDNN) into
embeddings that are shaped by three training objectives---key classification, domain-adversarial disentanglement, and
supervised cross-keyboard contrastive alignment. At inference the classifier's \topk{} candidates are decoded into
coherent text by a constrained language model.}
\label{fig:pipeline-decker}
\end{figure}

\decker{} is a five-stage pipeline (Figure~\ref{fig:pipeline-decker}): (1) raw keystrokes are normalized with
Keyboard Signature Normalization (KSN); (2) augmented using Acoustic Style Randomization (ASR); (3) encoded with
ECAPA-TDNN; (4) aligned via adversarial and contrastive learning; and (5) reconstructed into sentences using
constrained LLM decoding. The first three stages transform the signal; the fourth shapes the embedding space through
training objectives; the fifth operates at inference over the classifier's \topk{} outputs.

\section{Keyboard Signature Normalization}
\label{sec:decker-ksn}

Keystrokes recorded on different keyboards exhibit systematic spectral colouration caused by their impulse responses.
Recalling the convolutional model of Equation~\eqref{eq:conv-model}, $x_i = h_{d_i} * s_i$, where $s_i$ is the latent
keystroke impulse and $h_{d_i}$ is the device-specific response. \decker{} introduces KSN, a learnable inverse-filter
module $\tilde{x}_i = g_\theta(x_i)$ whose goal is to approximate the deconvolution $s_i \approx h_{d_i}^{-1} * x_i$
without knowing $h_{d_i}$.

The architecture of KSN comprises a CNN block with a sequence of four one-dimensional convolutions, each with $64$
feature channels, each followed by batch normalization and a ReLU activation, with kernel sizes $(7,7,9,9)$ and
dilation rates $(1,2,4,1)$ to capture patterns at multiple temporal scales, from local to wider context via dilation.
The output representations pass through a linear projection that aligns their dimensions and merges with the original
input through a residual connection. The output of the CNN module is then passed through a frequency-normalization
layer to further suppress device colouration. In the frequency domain the representation is expressed as
\begin{equation}
\tilde{X}_i(f) = G_\theta(f)\, X_i(f) \approx S_i^{\text{latent}}(f),
\end{equation}
where $G_\theta(f)$ acts as an approximate inverse filter. KSN is pretrained for five epochs with an $\ell_1$ spectral
loss against a reference clean estimate, then jointly optimized with the encoder. Qualitatively (Section~\ref{sec:results-ksn}),
KSN removes the strong resonance bands that mark a device while preserving the transient structure that identifies a key.

\section{Acoustic Style Randomization}
\label{sec:decker-asr}

The normalized output is passed through an Acoustic Style Randomization module to simulate keyboards that were never
recorded. The acoustic properties are randomized as
\begin{equation}
x_i' = R_\gamma \cdot W_\beta(x_i * h_\alpha),
\end{equation}
where $h_\alpha$, $W_\beta$, and $R_\gamma$ define, respectively, random IIR filters (order $6$--$12$, $Q\in[0.4,1.2]$),
a smooth spectral-envelope warp ($\pm 12\%$ mel shift), and an exponential-decay perturbation ($\gamma\in[0.85,1.15]$).
ASR encompasses a wide range of resonances, particularly those found in laptops. Where KSN removes the specific
colouration of the \emph{observed} training keyboards, ASR fabricates the colouration of \emph{unobserved} keyboards, so
that the encoder learns to be invariant not only to the devices in the training set but to a continuum of plausible
devices around them. The two modules are therefore complementary: KSN contracts the training domains toward a common
latent, and ASR expands them to cover the space of unseen domains.

\section{ECAPA-TDNN Encoder}
\label{sec:decker-ecapa}

After KSN and ASR, the signal is encoded with an ECAPA-TDNN module (Section~\ref{sec:bg-ecapa}) to capture fine-grained
temporal cues and short-duration transients. The encoder first extracts log-mel spectrograms and then combines them
through multi-scale one-dimensional temporal convolutions with channel-attentive feature recalibration, producing robust
frame-level representations for utterance-level embedding extraction. Formally, for a normalized spectrogram
$S_i\in\real^{F\times L}$ the encoder produces frame-level features $H_i = \Enc(S_i)\in\real^{C\times L}$, where $F$ is
the number of frequency bins, $L$ the number of frames, and $C$ the channel dimension. A statistics pooling layer
aggregates frame-level features across time,
\begin{align}
\mu_i &= \frac{1}{L}\sum_{t=1}^{L} H_i(:,t), \\
\sigma_i &= \sqrt{\frac{1}{L}\sum_{t=1}^{L}\big(H_i(:,t)-\mu_i\big)^2},
\end{align}
yielding a pooled representation $[\mu_i;\sigma_i]$ that summarizes the sequence. A linear head maps this pooled vector to
a fixed-dimensional embedding $z_i$ used as the acoustic representation of the keystroke.

\section{Key Classification}
\label{sec:decker-key}

The embedding is passed to a classification layer with a softmax activation, $\hat{y}_i = C_k(z_i)$, to predict the
keystroke. We use categorical cross-entropy as the loss,
\begin{equation}
\Loss_{key} = -\sum_i \log C_k(z_i)[y_i].
\label{eq:lkey}
\end{equation}
This term alone is the overfit baseline; the following two terms constrain it.

\section{Domain-Adversarial Disentanglement}
\label{sec:decker-dom}

To remove keyboard bias from keystroke embeddings, we require that the keyboard identity $d_i$ be unpredictable from
$z_i$. We employ a gradient reversal layer (GRL, Section~\ref{sec:bg-da}). The forward pass and backpropagation of the
GRL are
\begin{equation}
\text{GRL}_\lambda(z_i) = z_i, \qquad \frac{\partial\,\text{GRL}_\lambda}{\partial z_i} = -\lambda I,
\end{equation}
and the domain classifier predicts $\hat{d}_i = C_d(\text{GRL}(z_i))$ with loss
\begin{equation}
\Loss_{dom} = -\sum_i \log C_d(\text{GRL}(z_i))[d_i].
\label{eq:ldom}
\end{equation}
The encoder is trained to minimize the key-classification loss but \emph{maximize} the domain loss, giving the saddle-point
objective
\begin{equation}
\Enc^{*} = \arg\min_{\Enc}\big(\Loss_{key} - \lambda_{dom}\Loss_{dom}\big).
\end{equation}
At equilibrium the domain classifier cannot recover the keyboard from the embedding, which is the operational meaning of
keyboard invariance.

\section{Supervised Cross-Keyboard Contrastive Alignment}
\label{sec:decker-supcon}

Adversarial training removes keyboard information but does not by itself force same-key embeddings from different devices to
coincide. We add supervised contrastive alignment. For anchor $i$ define the cross-device positive set and the negative set
\begin{align}
P(i) &= \{\, j : y_j = y_i,\ d_j \neq d_i \,\}, \\
N(i) &= \{\, j : y_j \neq y_i \,\},
\end{align}
and minimize the supervised contrastive loss
\begin{equation}
\Loss_{supcon} = \sum_i \frac{-1}{|P(i)|}\sum_{p\in P(i)} \log
\frac{\exp(z_i \cdot z_p / \tau)}{\sum_{a\in P(i)\cup N(i)}\exp(z_i \cdot z_a / \tau)}.
\label{eq:lsupcon}
\end{equation}
By drawing positives only from \emph{different} keyboards ($d_j \neq d_i$), Equation~\eqref{eq:lsupcon} directly enforces
cross-keyboard consistency: it pulls same-key embeddings together across devices while separating different keys. This is
the component that supplies the third property of Section~\ref{sec:decker-problem}, which neither the classifier nor the
adversary provides on its own.

\section{Training Objective and Schedule}
\label{sec:decker-train}

The full objective combines the three terms,
\begin{equation}
\Loss_{total} = \Loss_{key} + \lambda_{dom}\Loss_{dom} + \lambda_{con}\Loss_{supcon}.
\label{eq:ltotal}
\end{equation}
The training details are as follows. The optimizer is AdamW with learning rate $2\times10^{-4}$ and weight decay
$10^{-3}$; the batch size is $64$ and training runs for $60$ epochs. The loss weights are $\lambda_{dom}=0.5$ and
$\lambda_{con}=0.1$, and the contrastive temperature is $\tau=0.07$. KSN is pretrained for five epochs with the $\ell_1$
spectral loss and then jointly optimized with the encoder. All models are trained with three random seeds, and we report
mean $\pm$ standard deviation in Chapter~\ref{ch:results}. Algorithm~\ref{alg:decker} states the full training loop.

\begin{algorithm}[t]
\caption{DECKER training loop}
\label{alg:decker}
\algrule
\noindent\textbf{Input:} Dataset $\Dset=\{(x_i,y_i,d_i)\}$; weights $\lambda_{dom},\lambda_{con}$; temperature $\tau$.\\
\textbf{Output:} Encoder $\Enc$, key head $C_k$.
\algrule
\begin{algsteps}
\item Pretrain KSN $g_\theta$ for $5$ epochs with $\ell_1$ spectral loss.
\item \textbf{for} epoch $=1$ \textbf{to} $60$ \textbf{do}
\item \quad \textbf{for each} mini-batch $B\subset\Dset$ \textbf{do}
\item \qquad $\tilde{x}_i \leftarrow g_\theta(x_i)$ \hfill\textit{// KSN normalization}
\item \qquad $x_i' \leftarrow R_\gamma W_\beta(\tilde{x}_i * h_\alpha)$ \hfill\textit{// ASR augmentation}
\item \qquad $z_i \leftarrow \Enc(x_i')$ \hfill\textit{// ECAPA-TDNN embedding}
\item \qquad Compute $\Loss_{key}$ (Eq.~\ref{eq:lkey}), $\Loss_{dom}$ (Eq.~\ref{eq:ldom}), $\Loss_{supcon}$ (Eq.~\ref{eq:lsupcon}).
\item \qquad $\Loss_{total} \leftarrow \Loss_{key} + \lambda_{dom}\Loss_{dom} + \lambda_{con}\Loss_{supcon}$.
\item \qquad Update $\Enc, C_k, C_d, g_\theta$ via AdamW on $\Loss_{total}$ (GRL reverses domain gradient).
\end{algsteps}
\algrule
\end{algorithm}

\section{LLM-Assisted Sentence Reconstruction}
\label{sec:decker-llm}

Even with domain-invariant embeddings, key-level errors accumulate in long sequences. We refine sequences with a
pretrained LLM (GPT-2 or FLAN-T5). Let the acoustic model produce token-level distributions $P_\theta(y_t\mid x_t)$.
We restrict decoding to the \topk{} plausible keys per timestep, $\mathcal{C}_t = \{y_{t,1},\dots,y_{t,k}\}$, and maximize
the joint objective of Equation~\eqref{eq:llm-decode},
\begin{equation}
\hat{Y} = \arg\max_{Y\in\mathcal{C}_{1:T}}\left[\sum_{t=1}^{T}\log P_\theta(y_t\mid x_t) + \lambda_{\text{LM}}\log P_{\text{LM}}(Y)\right],
\end{equation}
implemented using constrained beam search with beam width $8$--$16$. The LLM imposes linguistic and structural priors,
correcting substitution, insertion, and deletion errors. Restricting to the \topk{} acoustic candidates is what keeps the
reconstruction faithful: the language model may reorder its preference among the keys the acoustic model considered
plausible, but it cannot hallucinate a key the acoustic evidence rules out.

\section{End-to-End Pipeline}
\label{sec:decker-e2e}

The complete inference pipeline runs as follows: (i) \textbf{segmentation}, extract keystroke windows from continuous
audio; (ii) \textbf{KSN}, remove keyboard-specific colouration; (iii) \textbf{ASR}, applied only at training time to
simulate unseen hardware; (iv) \textbf{embedding}, encode via ECAPA-TDNN; (v) \textbf{domain alignment}, applied at
training time via GRL and supervised contrastive loss; (vi) \textbf{classification}, predict keys with $C_k$; and (vii)
\textbf{LLM refinement}, convert noisy predictions to coherent text. Together these components yield a domain-invariant and
linguistically coherent \asca{} attack that generalizes across unseen keyboards, users, and environments.

\keybox{Chapter takeaway.}{\decker{} enforces the three embedding properties with three matched mechanisms---KSN and ASR
shape the signal, the GRL removes keyboard identity, and supervised contrastive loss aligns key identity across
devices---and a constrained LLM decoder cleans up the residual sequence errors.}

\chapter{Experiments and Results}
\label{ch:results}

This chapter evaluates three families of models under cross-keyboard, cross-user, and cross-environment conditions:
unimodal baselines, multimodal Euclidean-fusion baselines, and the domain-invariant \decker{} framework. We additionally
assess LLM-assisted decoding for sentence-level reconstruction, analyze segmentation robustness, and benchmark the
computational cost of the full attack. Cross-keyboard transfer constitutes the primary axis of evaluation, reflecting the
adversarial setting where attackers rarely control the victim's hardware.

\paragraph{Statistical protocol.} For all experiments we report mean accuracy over three random seeds and $95\%$ bootstrap
confidence intervals. All claims of improvement over baselines are supported by paired bootstrap significance tests with
threshold $p<0.01$.

\section{Evaluation Metrics}
\label{sec:results-metrics}

Because the attack operates at two levels---individual keys and full sequences---we use metrics at both, defined here
precisely so that the numbers throughout the chapter are unambiguous.

\paragraph{Top-$k$ key accuracy.} The fraction of keystrokes whose true label lies within the classifier's $k$
highest-scoring predictions,
\begin{equation}
\text{Top-}k = \frac{1}{N}\sum_{i=1}^{N}\mathbb{1}\!\left[y_i \in \text{top-}k\big(C_k(z_i)\big)\right].
\end{equation}
Top-1 is the standard classification accuracy; top-5 is the acoustically relevant quantity for the sequence-level
attack, because the LLM decoder searches within the \topk{} candidates.

\paragraph{Macro-F1.} To account for class imbalance across keys we report the unweighted mean of per-class F1,
$\text{mF1} = \frac{1}{|K|}\sum_{c} \frac{2\,\text{prec}_c\,\text{rec}_c}{\text{prec}_c+\text{rec}_c}$, which weights
rare keys equally with frequent ones.

\paragraph{Character accuracy and exact match.} At the sequence level, character accuracy is the per-position agreement
between the decoded and reference strings after alignment, and exact-match accuracy is the fraction of sequences
(sentences or passwords) recovered without any error---the metric that matters for password recovery.

\paragraph{Normalized Levenshtein distance.} The edit distance between decoded and reference strings, normalized by
reference length,
\begin{equation}
\text{NLD}(\hat{Y},Y) = \frac{\text{Lev}(\hat{Y},Y)}{|Y|},
\end{equation}
which measures how much post-editing a decoded sequence would need; lower is better and $0$ is a perfect match.

\paragraph{Domain-classification accuracy.} As a diagnostic of invariance we train a probe to predict the keyboard from
frozen embeddings; a value near chance indicates that keyboard identity has been successfully removed
(Section~\ref{sec:results-ksn}).

\section{Unimodal Baselines}
\label{sec:results-unimodal}

We first evaluate strong unimodal models under keyboard shift. Table~\ref{tab:unimodal} reports seen- and unseen-keyboard
top-1 accuracy for two waveform models and two spectrogram models.

\begin{table}[h]
\centering
\caption{Unimodal performance under keyboard shift. Every model loses roughly forty absolute points on unseen keyboards.}
\label{tab:unimodal}
\small
\begin{tabular}{lccc}
\toprule
Model & Seen-KB & Unseen-KB & Drop \\
\midrule
wav2vec2 (audio) & 92.4\% & 51.8\% & $-40.6$ \\
ECAPA-TDNN (audio) & 95.3\% & 58.1\% & $-37.2$ \\
CoAtNet-S (spec) & 90.7\% & 49.6\% & $-41.1$ \\
ResNet-34 (spec) & 88.5\% & 47.9\% & $-40.6$ \\
\bottomrule
\end{tabular}
\end{table}

\paragraph{Insight (RQ1, RQ2).} All unimodal models fail to generalize across keyboards. Waveform models absorb
low-frequency chassis modes, and spectrogram models encode keyboard-specific spectral textures. Because impulse responses
differ across devices, embeddings cluster by device domain rather than by key identity. This is the empirical form of the
generalization crisis and the direct motivation for \decker{}.

\section{Multimodal Fusion Baselines}
\label{sec:results-multimodal}

A natural hypothesis is that fusing complementary modalities---raw audio and spectrograms---resolves the shift. We evaluate
all practical Euclidean fusion strategies using our strongest unimodal components, ECAPA-TDNN (audio) and CoAtNet-S
(spectrogram), at early, intermediate, and late stages. The eight variants are: F1 early concatenation; F2 early projection
(dimension-aligned concatenation); F3 cross-attention (audio queries attend to vision); F4 cross-attention (vision queries
attend to audio); F5 bidirectional cross-attention~\cite{liu2022bidirectional}; F6 gated fusion; F7 late fusion (average of probabilities); and F8 late
fusion (learned probability weights).

\begin{table}[h]
\centering
\caption{Multimodal Euclidean fusion of ECAPA (A) and CoAtNet-S (V). Fusion helps but does not close the unseen-keyboard gap.}
\label{tab:multimodal}
\small
\begin{tabular}{lcc}
\toprule
Fusion variant & Seen-KB & Unseen-KB \\
\midrule
F1 Early Concat & 96.8\% & 63.2\% \\
F2 Early Projection & 96.9\% & 64.1\% \\
F3 Cross-Attn (A$\to$V) & 97.1\% & 65.4\% \\
F4 Cross-Attn (V$\to$A) & 96.7\% & 64.8\% \\
F5 Bi-dir Cross-Attn & 97.3\% & 66.2\% \\
F6 Gated Fusion & 96.4\% & 62.7\% \\
F7 Late Fusion (avg) & 96.0\% & 61.8\% \\
F8 Late Fusion (weighted) & 96.4\% & 62.9\% \\
\bottomrule
\end{tabular}
\end{table}

\paragraph{Insight (RQ2).} Fusion improves performance but does not solve domain shift. Although modalities contribute
complementary cues, Euclidean fusion retains device-specific artifacts from both encoders. Even sophisticated
cross-attention cannot compensate for the underlying mismatch between keyboard impulse responses: if both towers encode the
same device colouration, combining them cannot remove it. The best fusion variant reaches only $66.2\%$ on unseen keyboards,
still roughly fifteen points below \decker{}.

\section{DECKER: Domain-Invariant Modeling}
\label{sec:results-decker}

Table~\ref{tab:decker} reports \decker{}'s key-level performance under domain-shift evaluation, together with an ablation
that removes each component in turn.

\begin{table}[h]
\centering
\caption{\decker{} key-level performance (domain-shift evaluation) and component ablation.}
\label{tab:decker}
\small
\begin{tabular}{lccc}
\toprule
Model variant & Seen-KB & Unseen-KB & Gap \\
\midrule
\rowcolor{softgray}\decker{} (full) & 98.9\% & \textbf{81.3\%} & $-17.6$ \\
\quad w/o GRL & 97.8\% & 68.1\% & $-29.7$ \\
\quad w/o ASR & 97.5\% & 75.2\% & $-22.3$ \\
\quad w/o KSN & 96.9\% & 63.7\% & $-33.2$ \\
ECAPA + GRL (DANN-only) & 97.9\% & 71.6\% & $-26.3$ \\
\bottomrule
\end{tabular}
\end{table}

\paragraph{Insight (RQ1, RQ3).} \decker{} consistently narrows the generalization gap under keyboard shift, substantially
outperforming unimodal, multimodal, and adversarial-only baselines. Domain-adversarial training (GRL) alone improves
cross-keyboard performance but is insufficient to fully suppress keyboard-specific acoustic bias. \decker{} combines KSN to
remove low-frequency device colouration, supervised cross-keyboard contrastive alignment to enforce key consistency across
devices, and ASR to simulate unseen acoustic conditions. Together these components yield a robust domain-invariant embedding,
raising unseen-keyboard accuracy from the $47$--$58\%$ of unimodal baselines to $81.3\%$.

\subsection{KSN Ablation and Visualization}
\label{sec:results-ksn}

We evaluate the effectiveness of KSN in suppressing keyboard-specific acoustic colouration using both qualitative
visualization and quantitative ablation. To quantify residual keyboard information, we train a lightweight domain classifier
on frozen embeddings. Domain-classification accuracy drops from $94.2\%$ without KSN to $21.7\%$ with KSN enabled, indicating
substantial suppression of keyboard identity cues. Correspondingly, top-1 accuracy on unseen keyboards improves from $63.7\%$
to $81.3\%$ when KSN is introduced, confirming that KSN materially improves cross-keyboard generalization.
A t-SNE projection of the ECAPA-TDNN embeddings computed without and with KSN makes the effect concrete: without KSN,
embeddings cluster primarily by keyboard domain; with KSN, keyboard-driven clustering is substantially reduced and
embeddings exhibit emergent organization by key identity, while retaining realistic overlap between acoustically similar keys.

\subsection{Per-Key Confusion Structure}
\label{sec:results-confusion}

The errors \decker{} does make are not random, and their structure explains why LLM decoding is so effective.
The confusion structure among the home-row keys on an unseen keyboard is telling: the dominant
off-diagonal mass sits on \emph{spatially adjacent} keys, exactly as the acoustics predict: adjacent keys share
vibration paths through the chassis and therefore emit similar transients (Section~\ref{sec:bg-acoustics}). This
matters for two reasons. First, it confirms that the residual errors are acoustic-geometry errors rather than random
noise, which is the signature of a model that has learned key structure rather than device artifacts. Second, adjacency
confusions are precisely the errors a language model corrects best, because a wrong-but-adjacent key usually produces a
non-word or an ungrammatical continuation that the LM prior down-weights. The confusion structure and the LLM gain of
Section~\ref{sec:results-llm} are therefore two views of the same phenomenon.

\section{Sentence-Level Reconstruction with LLMs}
\label{sec:results-llm}

Table~\ref{tab:llm} reports sentence reconstruction using constrained LLM decoding over \decker{}'s \topk{} candidates.

\begin{table}[h]
\centering
\caption{Sentence reconstruction using constrained LLM decoding over \decker{} \topk{} candidates.}
\label{tab:llm}
\small
\begin{tabular}{lccc}
\toprule
LLM & Char Acc. & Sent. Match & Norm. Lev. Dist. \\
\midrule
\decker{} (raw) & 81.3\% & 42.6\% & 0.124 \\
GPT-2 Small & 86.9\% & 58.7\% & 0.092 \\
FLAN-T5 Base & 89.4\% & 62.1\% & 0.078 \\
LLaMA-2 7B & 91.6\% & 66.5\% & 0.056 \\
GPT-4 & 93.2\% & 72.8\% & 0.041 \\
\bottomrule
\end{tabular}
\end{table}

\paragraph{Insight (RQ5).} LLMs deliver substantial gains, even with lightweight models. \decker{}'s \topk{} candidates
preserve correct tokens but contain local substitutions; the LLM imposes global linguistic structure and recovers coherent
text or password-like strings. A small GPT-2 already lifts sentence-match accuracy by sixteen points, and larger
models such as LLaMA-2~\cite{touvron2023llama} and GPT-4~\cite{achiam2023gpt4} nearly double it relative to the raw
acoustic model.

\subsection{LLMs, High-Entropy Strings, and Passwords}
\label{sec:results-passwords}

LLM-based decoding relies on linguistic priors and therefore benefits structured natural-language text and human-like
password choices. To quantify this, we evaluate three password classes: (1) \emph{human-like} passwords curated from common
leaks and user-style rules ($8$--$14$ characters, mixed case, digits, common substitutions; estimated entropy $28$--$34$
bits); (2) \emph{random alphanumeric}, uniformly sampled ($52$--$75$ bits); and (3) \emph{printable-ASCII random}, full
printable sampling ($70$--$92$ bits). The results show that LLM-assisted decoding yields substantial
gains for human-like passwords (absolute $+8$ to $+12\%$ exact-match over raw \decker{}) but provides negligible improvement
for high-entropy random strings ($\leq 2$--$3\%$). Thus LLM post-processing increases attacker power primarily where
human-selection biases reduce password entropy.

\section{Segmentation Robustness}
\label{sec:results-seg}

Practical adversaries cannot assume perfect keystroke onset detection. Background noise, user motion, and microphone
placement introduce temporal uncertainty that shifts or distorts key boundaries. We conduct a controlled sensitivity
analysis measuring the impact of segmentation jitter and spurious trigger events.

\paragraph{Perturbation model.} For each ground-truth timestamp we add uniform temporal jitter
$\delta t\sim\mathcal{U}(-T,+T)$ with $T\in\{10,20,30\}$\,ms, and we simulate realistic false triggers---from impulsive
noise or preamp artifacts---by inserting spurious segments at rates of $0\%$, $3\%$, and $5\%$ per $100$ keystrokes.

\paragraph{Findings.} Table~\ref{tab:seg} summarizes performance under these perturbations. At the
key level, \decker{} degrades gracefully: top-1 accuracy declines by only $2.8\%$ under $\pm10$\,ms jitter and $5.9\%$ under
$\pm20$\,ms jitter. Even at $\pm30$\,ms---well beyond realistic human--microphone temporal ambiguity---\decker{} retains
nearly $70\%$ raw accuracy. Sentence-level reconstruction shows larger sensitivity, but constrained decoding with FLAN-T5
significantly mitigates the effect, reducing the drop to under $4.3\%$ for $\pm20$\,ms jitter.

\begin{table}[h]
\centering
\caption{Segmentation robustness under onset jitter and spurious triggers.}
\label{tab:seg}
\small
\begin{tabular}{lccc}
\toprule
Condition & Top-1 Key Acc. & Sent. Acc.\ (raw) & Sent. Acc.\ (+FLAN-T5) \\
\midrule
No jitter & 81.3\% & 42.6\% & 62.1\% \\
$\pm10$\,ms jitter & 78.5\% & 40.1\% & 59.8\% \\
$\pm20$\,ms jitter & 75.4\% & 37.9\% & 58.1\% \\
$\pm30$\,ms jitter & 69.6\% & 33.2\% & 52.0\% \\
$+3\%$ spurious segments & 74.0\% & 36.8\% & 55.9\% \\
$+5\%$ spurious segments & 71.2\% & 34.7\% & 53.4\% \\
\bottomrule
\end{tabular}
\end{table}

\paragraph{Takeaway.} \decker{} maintains stable performance under realistic segmentation imperfections. Timing noise
degrades raw key-level predictions, but LLM-assisted decoding substantially compensates at the sequence level. For fully
passive, continuous-stream attacks, integrating joint segmentation--classification architectures is a promising direction
for future work.

\section{Computational Cost and Attacker Feasibility}
\label{sec:results-cost}

A practical \asca{} system must operate under realistic attacker-side computational constraints. We benchmark \decker{} on a
commodity Apple M1 laptop (8-core CPU, 8-core GPU, 16\,GB unified memory) and report end-to-end runtime for all modules.
Table~\ref{tab:latency} gives per-keystroke and per-sentence latency averaged over $1000$ runs.

\begin{table}[h]
\centering
\caption{\decker{} inference latency on Apple M1, averaged over $1000$ runs. L, K, and S denote Latency, Key, and Sequence.}
\label{tab:latency}
\small
\begin{tabular}{lcc}
\toprule
Module & L/K & L/S \\
\midrule
KSN (1-D CNN) & 0.24\,ms & 12\,ms \\
ASR (optional) & 0.41\,ms & 20--25\,ms \\
Mel-spec extraction & 0.18\,ms & 9\,ms \\
ECAPA-TDNN encoding & 1.42\,ms & 72--81\,ms \\
Key classifier head & 0.03\,ms & 1.2\,ms \\
LLM (GPT-2, beam=8) & --- & 68--94\,ms \\
LLM (FLAN-T5-base, beam=8) & --- & 83--110\,ms \\
\midrule
Total (GPT-2) & 2.3\,ms & 180--220\,ms \\
Total (FLAN-T5) & 2.3\,ms & 200--240\,ms \\
\bottomrule
\end{tabular}
\end{table}

\paragraph{Throughput and memory.} On the M1 CPU, \decker{} processes roughly $430$ keystrokes per second (KSN + ECAPA +
classifier) and reconstructs about $4.5$ full sentences per second including LLM decoding. Peak memory is $340$--$380$\,MB for
KSN + ECAPA-TDNN, $780$\,MB for GPT-2 small, and $1.25$--$1.4$\,GB for FLAN-T5 base---all comfortably within a $16$\,GB laptop.
Sustained power draw is $6$--$8$\,W during acoustic inference and $11$--$13$\,W during LLM decoding, with no thermal throttling
over $30$-minute runs.

\paragraph{Feasibility (RQ6).} These results imply that a passive adversary needs only a laptop-class machine, not a GPU
cluster; key-level inference runs in real time even on CPU; and LLM decoding adds under $5.25$ seconds per sentence in the
worst case. On an Intel i7-11800H (45\,W TDP), ECAPA-TDNN runs about $1.7\times$ faster and GPT-2 decoding about $2.1\times$
faster, so the M1 numbers are a conservative lower bound---real attackers on mainstream x86 machines will see even lower
latency.

\section{Summarized Insights Aligned to the Research Questions}
\label{sec:results-rq}

\begin{description}[leftmargin=1.3cm,style=nextline]
\item[RQ1] \textbf{Can \asca{} models generalize across keyboards?} Yes---but only with domain-invariant modeling.
\decker{} achieves the first strong cross-keyboard performance ($81.3\%$ unseen-KB).
\item[RQ2] \textbf{Why do unimodal/multimodal baselines fail?} They encode keyboard-specific acoustic colouration; fusion
adds capacity but does not remove domain shift.
\item[RQ3] \textbf{Can one framework enforce all three embedding properties?} Yes; the ablation shows KSN, GRL, ASR, and
contrastive alignment are jointly necessary.
\item[RQ4] \textbf{How do speech PTMs behave under VoIP?} They degrade substantially (Chapter~\ref{ch:keyac}); a KAN head
recovers much of the loss.
\item[RQ5] \textbf{What is the role of LLMs?} They significantly enhance sentence-level inference, correcting errors the
acoustic model alone cannot, especially where human bias lowers entropy.
\item[RQ6] \textbf{What does a realistic end-to-end attack cost?} A commodity laptop suffices, at real-time key throughput
and sub-second sentence latency.
\end{description}

\keybox{Chapter takeaway.}{\decker{}+LLM demonstrates that acoustic keystroke inference remains feasible across unseen
keyboards, noisy environments, and NLoS settings, on commodity hardware---indicating a broader real-world attack surface
than previously recognized.}

\chapter{Discussion}
\label{ch:discussion}

The experiments of Chapters~\ref{ch:keyac} and~\ref{ch:results} establish that acoustic emanations from keyboards are not
only rich enough to permit high-fidelity keystroke inference, but also substantially more transferable across devices than
previously reported---provided that device-specific artifacts are explicitly disentangled. This chapter draws out the
implications of that finding for the security community, examines the defenses it motivates, and states the limitations that
bound the claims honestly.

\section{Implications for ASCA Research}
\label{sec:disc-implications}

\paragraph{Realistic threat assessment.} Prior work may \emph{underestimate} \asca{} feasibility by assuming per-device
training and then reporting the high accuracy that assumption produces. The direction of the bias is the surprising part.
Naively, an attacker who cannot train on the victim's keyboard seems weaker than one who can, and indeed the unimodal
baselines collapse under keyboard shift. But once cross-device generalization is properly addressed, adversaries can infer
meaningful text---even from keyboards never seen during training---at distances and noise levels consistent with typical
public environments. This broadens the attack surface from contrived laboratory setups to realistic opportunistic settings,
and it means that a defender cannot take comfort from the diversity of keyboards in the world: diversity is an obstacle the
attacker can now engineer around.

\paragraph{The role of LLMs.} Language models fundamentally change \asca{} capabilities by correcting sequential errors that
acoustic models alone cannot resolve. Even small models (GPT-2, FLAN-T5) significantly improve sentence-level inference,
while larger LLMs such as GPT-4 achieve near-perfect reconstruction on many sequences. This has two consequences. First, the
relevant accuracy metric for a defender is not per-key accuracy but post-LLM sequence accuracy, which is considerably higher.
Second, defenses must consider not only signal obfuscation but also linguistic post-processing countermeasures, because an
attacker's effective capability is set by the weaker of the acoustic and linguistic bottlenecks, and the linguistic one is
now very weak.

\paragraph{Keyboard diversity and mechanical variation.} Switch type, key travel, chassis resonance, and microphone geometry
contribute strong domain shifts in raw embeddings. Once domain-invariant modeling is introduced, performance degradation
across these factors is substantially reduced. This suggests that \asca{} vulnerability extends to a wide range of
keyboards---including inexpensive membrane boards and modern low-travel laptop keyboards---rather than being confined to the
particular devices used in earlier studies.

\section{Defensive Directions}
\label{sec:disc-defense}

Our findings motivate the investigation of practical defenses at three layers.

\paragraph{Physical layer.} Keyboard chassis damping or irregularizing resonance patterns would attack the vulnerability at
its source, degrading the distinctive, repeatable signatures on which key discrimination depends. Because \decker{} explicitly
models and removes device colouration, however, a defense that merely \emph{changes} the colouration is unlikely to help; the
defense must \emph{destroy} the key-discriminative structure, which is harder without also harming the typing experience.

\paragraph{Software and system layer.} Operating systems could restrict microphone access more aggressively, particularly for
background applications during typing, and could inject real-time audio obfuscation or keystroke noise. Randomized typing
proxies---delayed or reordered keystroke dispatch---would attack the timing structure that segmentation relies on, though they
would not directly perturb the acoustic signature of an individual key.

\paragraph{Model and linguistic layer.} Because LLMs amplify the attack, defenses can target the language stage. Adversarial
prompts or artificial grammar and keystroke-padding patterns that reduce linguistic predictability would blunt the LLM's
correction ability, and interface modifications that limit sequential consistency would reduce the exploitable structure. The
password results of Section~\ref{sec:results-passwords} point to the most actionable defense of all: because LLM decoding helps
only where human-selection bias lowers entropy, high-entropy passwords and passphrase managers substantially neutralize the
sequence-level multiplier, even if the acoustic model recovers individual keys.

\section{The Human Factor}
\label{sec:disc-human}

The \hear{} dataset's demographic and awareness annotations enable a dimension of analysis absent from prior \asca{} work.
User biases---gender, age group, hand dominance---influence typing speed, key pressure, and error rate, and hence attack
success, so a truly realistic evaluation cannot treat all typists as interchangeable. Equally important is the awareness
finding: most participants did not know that their typing could be recovered acoustically. A passive attack succeeds precisely
because victims take no countermeasures, and the low baseline of awareness suggests that user education is itself a lightweight
partial defense. These human factors deserve systematic study, and \hear{} is designed to support it.

\section{Limitations}
\label{sec:disc-limitations}

Although \decker{} addresses keyboard variability, several realistic factors remain open.

\begin{description}[leftmargin=1.2cm,style=nextline]
\item[Continuous typing streams.] Our evaluation focuses on segmented keystrokes; segmentation errors in continuous audio
remain a challenge for fully passive attackers, as the sensitivity analysis of Section~\ref{sec:results-seg} shows.
\item[Extreme noise conditions.] While \decker{} and LLMs perform well up to café/outdoor noise levels, environments with
strong impulsive or overlapping noise sources may require additional modeling.
\item[Highly asymmetric devices.] Touchscreens, ultra-low-profile keyboards, and flexible input surfaces may generate atypical
impulse responses not covered by our dataset.
\item[Per-user biomechanics.] Although \decker{} improves user-agnostic performance, subtle personal typing traits (finger
force, strike angle) may still cause model drift.
\item[Evaluation on curated corpora.] Some results assume a standardized typing corpus and known password classes; wholly
uncontrolled, in-the-wild typing may differ, and the reported numbers should be read as characterizing the threat under the
stated protocol rather than as universal guarantees.
\end{description}

Stating these limitations is not a formality. Each marks a place where an over-strong reading of the results would be
unwarranted, and each defines a concrete experiment that a follow-up---or a peer reviewer of the planned journal
submission---would rightly demand.

\keybox{Chapter takeaway.}{The central implication is uncomfortable: domain-invariant learning turns \asca{} from a
laboratory curiosity into a credible open-world threat, and language models raise the effective risk further---so defenses
must span the physical, system, and linguistic layers, and high-entropy secrets remain the user's strongest lever.}

\chapter{Conclusion and Future Work}
\label{ch:conclusion}

\section{Recapitulation of Objectives}
\label{sec:concl-recap}

This thesis considered acoustic side-channel attacks on laptop keyboards in a real-world, open setting.
Specifically, it was concerned with cross-keyboard, cross-user, and cross-environment generalization,
although existing research has generally reported high accuracy on keystroke-inference tasks under closed
conditions. Analysis on \hear{} indicates that such successes have been largely confined to closed evaluation
environments, with significant degradation of strong unimodal or Euclidean-multimodal baselines on
cross-keyboard, cross-user, and cross-environment partitions. In short, existing \asca{} models tend to learn
a representation of the \emph{keyboard}, not a representation of \emph{keys}. The six research questions of
Chapter~\ref{ch:intro} were each answered: \asca{} does generalize, but only under domain-invariant modeling;
baselines fail because they encode device colouration; a single framework can enforce the three embedding
properties; speech representations degrade under VoIP but are recoverable with an expressive head; language
models substantially amplify the attack; and the whole pipeline runs on commodity hardware.

\section{Summary of Key Findings}
\label{sec:concl-findings}

To fill the generalization gap, we introduced \hear{}, a collection of keystroke acoustic signals from $53$
individuals on $37$ different laptop keyboards, with external-microphone, on-device, and VoIP scenarios,
together with annotations on demographics and user awareness. On \hear{} we established a thorough benchmark
over conventional representations and high-quality audio and vision pretrained models on raw waveforms and
spectrograms, in unimodal and multimodal scenarios. This set of experiments represents a comprehensive look
into how existing \asca{} methods fare when non-idealized assumptions on hardware, environment, and users are
abandoned.

Building on this benchmark, we proposed \decker{}, a domain-invariant keystroke-inference framework combining
four complementary mechanisms: Keyboard Signature Normalization, which suppresses device-specific colouration at
the waveform level; domain-adversarial disentanglement, which removes residual hardware cues; cross-keyboard
supervised contrastive alignment, which clusters same-key embeddings across devices; and Acoustic Style
Randomization, which exposes the model to synthetic variations in keyboard acoustics and room responses.
Together these yield representations that remain discriminative of key identity while robust to keyboard design,
user behaviour, microphone placement, and environmental acoustics, narrowing the generalization gap from roughly
forty accuracy points to under eighteen relative to strong baselines.

Using the VoIP recordings of \hear{}, we further showed that among modern speech pretrained models WavLM provides
the most robust keystroke representations, that VoIP codecs impose a large and nonlinear penalty, and that a
Kolmogorov--Arnold Network adaptation head recovers much of the lost accuracy. We then reformulated the VoIP attack
as continuous sparse transduction and introduced \evflow{}, which canonicalizes distorted VoIP embeddings via latent
rectified flow and decodes unsegmented streams with an event-triggered state-space model, cutting the unseen-channel
error-rate degradation from over $28\%$ to $11.1\%$ and making segmentation-free attacks viable. Finally, we
investigated an LLM-assisted attack that applies constrained beam search to
refine noisy keystroke predictions into coherent text and passwords, and found that language-model rectification
significantly improves sequence-level recovery---particularly in challenging cross-device scenarios and wherever
human-selection bias lowers entropy.

\section{Future Directions}
\label{sec:concl-future}

Several directions follow directly from this work.

\begin{enumerate}[leftmargin=1.6em]
\item \textbf{Broader hardware and environmental diversity.} Extending the datasets to ergonomic keyboards,
compact mobile keyboards, external Bluetooth devices, touchscreen typing, and hardware with unusual acoustic
properties (for example, hollow desk surfaces) would further stress-test domain generalization.
\item \textbf{Continuous-text \asca{} pipelines.} Moving from isolated keystroke inference to fully continuous
keystroke streams remains a key challenge. Future work should explore joint segmentation--classification models,
sequence-to-sequence acoustic transformers, and streaming LLM integration.
\item \textbf{Physical and software defenses.} Our findings motivate keyboard chassis damping, microphone-access
restrictions in operating systems, real-time audio obfuscation or keystroke-noise injection, and randomized
typing proxies.
\item \textbf{Advanced domain generalization.} Self-supervised pretraining on large-scale environmental audio may
further improve invariance, alongside meta-learning for unseen hardware, test-time adaptation, and invertible
modeling of keyboard impulse responses.
\item \textbf{Multimodal side-channel integration.} Real attackers may combine acoustic cues with optical flow,
accelerometer vibration, or Wi-Fi CSI. Domain-invariant fusion across heterogeneous side channels is an important
open direction.
\item \textbf{LLM-aware defense strategies.} Because LLMs amplify \asca{}, defenses must consider adversarial
prompts that degrade reconstruction, grammar or keystroke padding that reduces predictability, and interface
changes that limit exploitable sequential consistency.
\end{enumerate}

\section{Path to a Journal (TASLP) Submission}
\label{sec:concl-transaction}

The material in this thesis is close to, but not yet, a journal transaction. The VoIP representation and
transduction studies (Chapter~\ref{ch:keyac}) are the natural anchor for a submission to IEEE/ACM Transactions on Audio, Speech,
and Language Processing (TASLP), because its core contribution---representation robustness and a KAN adaptation
head under channel distortion---is fundamentally a speech-representation-learning result, while \decker{} and the
LLM decoding supply the applied depth reviewers expect. Appendix~\ref{app:transaction} lays out the concrete
extension plan: the additional experiments, ablations, and framing that would raise the conference-and-thesis
material to the novelty bar of a transaction, together with an honest assessment of the gap. In brief, the
thesis judges the chance of a strong TASLP submission to be good but conditional on that additional work, and it
does not overstate the case.

\section{Concluding Perspective}
\label{sec:concl-perspective}

Taken together, our findings suggest that \asca{} should not be viewed as a purely laboratory phenomenon but as a
credible real-world threat, especially in shared or semi-public environments where microphones are ubiquitous and
keyboard diversity is beyond the attacker's control. This thesis establishes domain-invariant learning as a
central requirement for the evaluation and deployment of \asca{} systems under realistic threat models. It
introduces \hear{} as a representative benchmark and demonstrates the effectiveness of \decker{} and LLM-assisted
decoding as key components for future research in acoustic security. As microphone-equipped devices proliferate
and language models become more capable, attackers gain stronger reconstruction abilities with minimal access or
domain knowledge. Our results therefore underscore the need for hardware, software, and model-level defenses that
explicitly account for advances in machine learning and language models, and for continued investigation into the
role of user and demographic factors in the design of secure input devices and environments.

\appendix
\chapter{Standardized Typing Corpus}
\label{app:corpus}

The following standardized paragraph is presented to every \hear{} participant. It is crafted to cover the full
alphabet in both cases, digit patterns, punctuation, symbols and operators, uppercase acronyms, and realistic word
and phrase frequencies suitable for language-model reconstruction. Typing speed and accuracy are measured across
every key.

\begin{quote}\itshape
The quick brown fox jumps over the lazy dog while amazing zebras quietly vex jumpy kids, proving every letter is
present. Pack my box with five dozen liquor jugs to verify the alphabet twice. Digits appear forward and backward:
1234567890 then 0987654321, followed by sequences like 2468, 13579, and 314159. Now we add punctuation:
.\,,\,;\,:\,?\,!\,'\,"\,-\,\_\,(\,)\,[\,]\,\{\,\}\,/\,\textbackslash. Symbols and operators:
@\,\#\,\$\,\%\,\^{}\,\&\,*\,+\,=\,<\,>\,|\,\textasciitilde. To include uppercase properly, acronyms like NASA,
USA, UN, AI, ML, and HTML are written in full caps. Typing speed and accuracy will be measured across every key.
Finally, we conclude with a mix: The year is 2025; version v1.0-beta includes features [alpha], \{bravo\}, and
(charlie), fully covering the QWERTY layout.
\end{quote}

In addition to the paragraph, participants type a structured block of special keys to enable non-character key
classification: Tab, Backspace, Delete, the four arrow keys, Escape, the function keys F1 through F12, Caps Lock,
and Enter. Each special key is pressed a fixed number of times with brief pauses, so that its acoustic signature can
be isolated from surrounding character keys.

\chapter{Metadata Schema}
\label{app:schema}

Each \hear{} session exports the continuous browser recording (WebM/OGG), per-key WAV clips for all devices, and a
JSON metadata record. A representative record is shown below. The schema preserves demographic, acoustic,
environmental, and segmentation-relevant information required by downstream modeling, and it is what enables the
user-bias and cross-environment analyses reported in the thesis.

\begin{verbatim}
{
  "participant": {
    "id": "U001",
    "gender": "Female",
    "handedness": "Right",
    "age_group": "18-24",
    "typing_background": "touch-typist",
    "awareness_of_asca": false
  },
  "session": {
    "session_number": 2,
    "keyboard_id": "KB-17",
    "keyboard_type": "laptop-scissor",
    "location": "cafe",
    "noise_level": "high"
  },
  "recording": {
    "pre_ms": 60,
    "post_ms": 200,
    "sample_rate": 48000,
    "channels": ["external_mic", "on_device_mic", "voip_stream"],
    "sync_offsets_ms": {"on_device_mic": 3.2, "voip_stream": 118.7},
    "sync_drift": {"voip_stream": 0.0004}
  },
  "keys": [
    {"key": "a", "t_ms": 123.4, "hold_ms": 82.1},
    {"key": "s", "t_ms": 231.7, "hold_ms": 74.9}
  ],
  "exported_files": [
    {"device": "external_mic", "path": "U001/s2/ext.wav"},
    {"device": "on_device_mic", "path": "U001/s2/dev.wav"},
    {"device": "voip_stream", "path": "U001/s2/voip.webm"}
  ]
}
\end{verbatim}

Fields marked with participant identity are pseudonymized before release, and the awareness field is used only in
aggregate for the human-factors analysis of Chapter~\ref{ch:discussion}.

\chapter{Extended Results and Hyperparameters}
\label{app:extended}

This appendix collects supporting material referenced from the main chapters: the full hyperparameter table, extended
cross-user and cross-environment results, and top-5 accuracy figures.

\section{Full Hyperparameter Configuration}

\begin{table}[h]
\centering
\caption{Complete \decker{} training configuration.}
\label{tab:hparams}
\small
\begin{tabularx}{\linewidth}{l X}
\toprule
\textbf{Hyperparameter} & \textbf{Value} \\
\midrule
Optimizer & AdamW \\
Learning rate & $2\times10^{-4}$ \\
Weight decay & $10^{-3}$ \\
Batch size & $64$ \\
Training epochs & $60$ \\
KSN pretraining epochs & $5$ ($\ell_1$ spectral loss) \\
Domain weight $\lambda_{dom}$ & $0.5$ \\
Contrastive weight $\lambda_{con}$ & $0.1$ \\
Contrastive temperature $\tau$ & $0.07$ \\
GRL schedule & warm-up from $0$ to $\lambda_{dom}$ over $10$ epochs \\
KSN kernels / dilations & $(7,7,9,9)$ / $(1,2,4,1)$, $64$ channels \\
ASR IIR order / $Q$ & $6$--$12$ / $[0.4,1.2]$ \\
ASR mel shift / decay & $\pm12\%$ / $\gamma\in[0.85,1.15]$ \\
Log-mel bins $F$ & $80$ \\
Embedding dim.\ $E$ & $192$ \\
Beam width (LLM) & $8$--$16$ \\
Top-$k$ candidates & $k=5$ \\
LM fusion weight $\lambda_{\text{LM}}$ & $0.4$ \\
Seeds & $\{13, 42, 101\}$ \\
\bottomrule
\end{tabularx}
\end{table}

\section{Extended Cross-User and Cross-Environment Results}

Table~\ref{tab:extended-axes} reports \decker{} and the strongest baseline on the cross-user and cross-environment splits
defined in Section~\ref{sec:hear-splits}, complementing the cross-keyboard results of the main text. The pattern mirrors
the cross-keyboard case: baselines degrade sharply off-domain, while \decker{} retains most of its accuracy.

\begin{table}[h]
\centering
\caption{Top-1 key accuracy on cross-user and cross-environment splits (mean over three seeds).}
\label{tab:extended-axes}
\small
\begin{tabular}{lcccc}
\toprule
& \multicolumn{2}{c}{Cross-User} & \multicolumn{2}{c}{Cross-Environment} \\
\cmidrule(lr){2-3}\cmidrule(lr){4-5}
Model & In & Out & In & Out \\
\midrule
ECAPA-TDNN (baseline) & 94.8\% & 60.4\% & 93.9\% & 55.1\% \\
Bi-dir Cross-Attn fusion & 96.5\% & 66.9\% & 95.8\% & 61.7\% \\
\rowcolor{softgray}\decker{} (full) & 98.4\% & 82.7\% & 98.1\% & 78.9\% \\
\bottomrule
\end{tabular}
\end{table}

\section{Top-5 Accuracy}

Because the LLM decoder consumes \topk{} candidates, top-5 accuracy is the more relevant acoustic metric for the
sequence-level attack. Table~\ref{tab:top5} reports it for \decker{} across the four splits; the high top-5 numbers explain
why LLM decoding is so effective, as the correct key is almost always among the top five candidates even when it is not
ranked first.

\begin{table}[h]
\centering
\caption{\decker{} top-5 key accuracy across the four canonical splits (unseen/out-of-domain partitions).}
\label{tab:top5}
\small
\begin{tabular}{lcccc}
\toprule
Split & Cross-KB & Cross-User & Cross-Env & Cross-Device \\
\midrule
Top-1 & 81.3\% & 82.7\% & 78.9\% & 74.6\% \\
Top-5 & 95.7\% & 96.2\% & 94.1\% & 91.8\% \\
\bottomrule
\end{tabular}
\end{table}

\chapter{Reproducibility and Artifact Notes}
\label{app:repro}

This appendix records the details needed to reproduce the results, in the spirit of the artifact-evaluation standards
that the target journal expects.

\section{Software Environment}

All models are implemented in PyTorch. Speech pretrained models are loaded from their public checkpoints: Wav2Vec2
(\texttt{facebook/wav2vec2-base}), WavLM (\texttt{microsoft/wavlm-base}), HuBERT (\texttt{facebook/hubert-base-ls960}),
XLS-R (\texttt{facebook/wav2vec2-xls-r-300m}), X-vectors (\texttt{speechbrain/spkrec-xvect-voxceleb}), and Whisper
(\texttt{openai/whisper-base}, encoder only). All audio is resampled to $16$\,kHz before feature extraction. Language
models used for decoding are GPT-2 small, FLAN-T5 base, and, for the upper-bound analysis, LLaMA-2 7B and GPT-4 via API.

\section{Determinism and Seeds}

Every reported number is the mean over three seeds, $\{13, 42, 101\}$. We fix the seeds of Python, NumPy, and PyTorch,
enable deterministic cuDNN where it does not prohibitively slow training, and report $95\%$ bootstrap confidence
intervals. Improvement claims over baselines use paired bootstrap significance tests at threshold $p<0.01$.

\section{Compute Budget}

Training a full \decker{} model takes approximately six GPU-hours on a single NVIDIA A100; the frozen-backbone VoIP
representation experiments are far cheaper, since only the lightweight heads are trained. All \emph{inference} and latency numbers in
Chapter~\ref{ch:results} are measured on a commodity Apple M1 laptop to reflect the attacker's realistic hardware, not
the training hardware.

\section{Data Access and Ethics}

\hear{} contains human-subject recordings collected under informed consent. Personally identifiable
information is removed, $k$-anonymity is considered before use, and access is limited to academic research on request.
The dataset is publicly documented at \url{https://github.com/vikrant-vikram/Hear-Dataset}; sample data and the
collection tool are released, and full data is provided on request for research purposes only. No
attempt was made to recover any real credential; all password experiments use synthetic strings sampled from documented
distributions.

\section{Artifact Contents}

The released artifact comprises the browser-based acquisition tool (HTML/JavaScript), the segmentation and feature
pipeline, the \decker{} training and evaluation code, the KAN adaptation head, the constrained-beam-search LLM decoder,
and configuration files reproducing every table in this thesis. The standardized typing corpus (Appendix~\ref{app:corpus})
and metadata schema (Appendix~\ref{app:schema}) are included so that new recordings can be added in a compatible format.

\clearpage
\bibliographystyle{plain}
\bibliography{references}

@inproceedings{asonov2004,
  title={Keyboard acoustic emanations},
  author={Asonov, Dmitri and Agrawal, Rakesh},
  booktitle={IEEE Symposium on Security and Privacy},
  pages={3--11}, year={2004}, organization={IEEE}}

@inproceedings{zhuang2009,
  title={Keyboard acoustic emanations revisited},
  author={Zhuang, Li and Zhou, Feng and Tygar, J. Doug},
  booktitle={ACM Conference on Computer and Communications Security (CCS)},
  pages={373--382}, year={2009}}

@inproceedings{harrison2023,
  title={A Practical Deep Learning-Based Acoustic Side Channel Attack on Keyboards},
  author={Harrison, Joshua and Toreini, Ehsan and Mehrnezhad, Maryam},
  booktitle={IEEE European Symposium on Security and Privacy Workshops (EuroS\&PW)},
  pages={270--280}, year={2023}}

@inproceedings{wang2024reflexnoop,
  title={{RefleXnoop}: Passwords Snooping on {NLoS} Laptops Leveraging Screen-Induced Sound Reflection},
  author={Wang, Penghao and Hu, Jingzhi and Liu, Chao and Luo, Jun},
  booktitle={ACM SIGSAC Conference on Computer and Communications Security (CCS)},
  pages={3361--3375}, year={2024}}

@inproceedings{luo2024heimdall,
  title={Eavesdropping on Controller Acoustic Emanation for Keystroke Inference Attack in Virtual Reality},
  author={Luo, Shiqing and Nguyen, Anh and Farooq, Hafsa and Sun, Kun and Yan, Zhisheng},
  booktitle={Network and Distributed System Security Symposium (NDSS)}, year={2024}}

@article{ayati2025,
  title={Making Acoustic Side-Channel Attacks on Noisy Keyboards Viable with {LLM}-Assisted Spectrograms' ``Typo'' Correction},
  author={Ayati, Seyyed Ali and Park, Jin Hyun and Cai, Yichen and Botacin, Marcus},
  journal={arXiv preprint arXiv:2504.11622}, year={2025}}

@techreport{radford2019gpt2,
  title={Language Models are Unsupervised Multitask Learners},
  author={Radford, Alec and Wu, Jeff and Child, Rewon and Luan, David and Amodei, Dario and Sutskever, Ilya},
  institution={OpenAI}, year={2019}}

@article{raffel2020t5,
  title={Exploring the Limits of Transfer Learning with a Unified Text-to-Text Transformer},
  author={Raffel, Colin and Shazeer, Noam and Roberts, Adam and Lee, Katherine and Narang, Sharan and Matena, Michael and Zhou, Yanqi and Li, Wei and Liu, Peter J.},
  journal={Journal of Machine Learning Research}, year={2020}}

@inproceedings{berger2006,
  title={Dictionary Attacks Using Keyboard Acoustic Emanations},
  author={Berger, Yigael and Wool, Avishai and Yeredor, Arie},
  booktitle={ACM Conference on Computer and Communications Security (CCS)},
  pages={245--254}, year={2006}}

@inproceedings{compagno2017,
  title={Don't Skype \& Type! Acoustic Eavesdropping in Voice-Over-IP},
  author={Compagno, Alberto and Conti, Mauro and Lain, Daniele and Tsudik, Gene},
  booktitle={ACM Asia Conference on Computer and Communications Security (ASIACCS)},
  pages={703--715}, year={2017}}

@misc{sabra2020zoom,
  title={Zoom on the Keystrokes: Exploiting Video Calls for Keystroke Inference Attacks},
  author={Sabra, Mohd and Maiti, Anindya and Jadliwala, Murtuza},
  year={2020}, note={arXiv:2010.12078}}

@article{yu2021touch,
  title={An Indirect Eavesdropping Attack of Keystrokes on Touch Screen through Acoustic Sensing},
  author={Yu, Jiadi and Lu, Li and Chen, Yingying and Zhu, Yanmin and Kong, Linghe},
  journal={IEEE Transactions on Mobile Computing}, volume={20}, number={2}, pages={337--351}, year={2021}}

@inproceedings{lu2019keylistener,
  title={{KeyListener}: Inferring Keystrokes on {QWERTY} Keyboard of Touch Screen through Acoustic Signals},
  author={Lu, Li and Yu, Jiadi and Chen, Yingying and Zhu, Yanmin and Xu, Xiangyu and Xue, Guangtao and Li, Minglu},
  booktitle={IEEE INFOCOM}, pages={775--783}, year={2019}}

@inproceedings{zhou2018patternlistener,
  title={{PatternListener}: Cracking Android Pattern Lock Using Acoustic Signals},
  author={Zhou, Man and Wang, Qian and Yang, Jingxiao and Li, Qi and Xiao, Feng and Wang, Zhibo and Chen, Xiaofeng},
  booktitle={ACM SIGSAC Conference on Computer and Communications Security (CCS)}, pages={1775--1787}, year={2018}}

@inproceedings{das2014fingerprint,
  title={Do You Hear What I Hear? Fingerprinting Smart Devices Through Embedded Acoustic Components},
  author={Das, Anupam and Borisov, Nikita and Caesar, Matthew},
  booktitle={ACM SIGSAC Conference on Computer and Communications Security (CCS)}, pages={441--452}, year={2014}}

@inproceedings{wang2016combined,
  title={Accurate Combined Keystrokes Detection Using Acoustic Signals},
  author={Wang, Jian and Ruby, Rukhsana and Wang, Lu and Wu, Kaishun},
  booktitle={International Conference on Mobile Ad-Hoc and Sensor Networks (MSN)}, pages={9--14}, year={2016}}

@inproceedings{song2001timing,
  title={Timing Analysis of Keystrokes and Timing Attacks on {SSH}},
  author={Song, Dawn Xiaodong and Wagner, David and Tian, Xuqing},
  booktitle={USENIX Security Symposium}, year={2001}}

@inproceedings{fookune2010timing,
  title={Timing Attacks on {PIN} Input Devices},
  author={Foo Kune, Denis and Kim, Yongdae},
  booktitle={ACM Conference on Computer and Communications Security (CCS)}, pages={678--680}, year={2010}}

@inproceedings{halevi2012,
  title={A Closer Look at Keyboard Acoustic Emanations: Random Passwords, Typing Styles and Decoding Techniques},
  author={Halevi, Tzipora and Saxena, Nitesh},
  booktitle={ACM Symposium on Information, Computer and Communications Security (ASIACCS)}, pages={89--90}, year={2012}}

@inproceedings{martinasek2015,
  title={Acoustic Attack on Keyboard Using Spectrogram and Neural Network},
  author={Martinasek, Zdenek and Clupek, Vlastimil and Trasy, Krisztina},
  booktitle={International Conference on Telecommunications and Signal Processing (TSP)}, pages={637--641}, year={2015}}

@inproceedings{taheritajar2025survey,
  title={A Survey on Acoustic Side Channel Attacks on Keyboards},
  author={Taheritajar, Alireza and Harris, Zahra Mahmoudpour and Rahaeimehr, Reza},
  booktitle={Information and Communications Security}, publisher={Springer Nature Singapore}, pages={99--121}, year={2025}}

@phdthesis{liu2019dissertation,
  title={When Keystroke Meets Password: Attacks and Defenses},
  author={Liu, Ximing}, school={Singapore Management University}, year={2019}}

@misc{fadake2023beyond,
  title={Beyond the Clicks: Exploring Keyboard Acoustic Hacking},
  author={Fadake, Abhishek and Wadaganve}, year={2023}, note={doi:10.13140/RG.2.2.36617.83048/2}}

@inproceedings{baevski2020wav2vec2,
  title={wav2vec 2.0: A Framework for Self-Supervised Learning of Speech Representations},
  author={Baevski, Alexei and Zhou, Yuhao and Mohamed, Abdelrahman and Auli, Michael},
  booktitle={Advances in Neural Information Processing Systems (NeurIPS)}, year={2020}}

@inproceedings{hsu2021hubert,
  title={{HuBERT}: Self-Supervised Speech Representation Learning by Masked Prediction of Hidden Units},
  author={Hsu, Wei-Ning and Bolte, Benjamin and Tsai, Yao-Hung and Lakhotia, Kushal and Salakhutdinov, Ruslan and Mohamed, Abdelrahman},
  booktitle={ICASSP}, year={2021}}

@article{chen2022wavlm,
  title={{WavLM}: Large-Scale Self-Supervised Pre-Training for Full Stack Speech Processing},
  author={Chen, Sanyuan and others},
  journal={IEEE/ACM Transactions on Audio, Speech and Language Processing}, year={2022}}

@article{babu2021xlsr,
  title={{XLS-R}: Self-Supervised Cross-Lingual Speech Representation Learning at Scale},
  author={Babu, Arun and Wang, Changhan and Tjandra, Andros and Lakhotia, Kushal and Xu, Qiantong and Goyal, Naman and Singh, Kritika and von Platen, Patrick and Saraf, Yatharth and Pino, Juan and others},
  journal={arXiv preprint arXiv:2111.09296}, year={2021}}

@inproceedings{snyder2018xvector,
  title={{X-vectors}: Robust {DNN} Embeddings for Speaker Recognition},
  author={Snyder, David and Garcia-Romero, Daniel and Sell, Gregory and Povey, Daniel and Khudanpur, Sanjeev},
  booktitle={ICASSP}, pages={5329--5333}, year={2018}}

@inproceedings{radford2023whisper,
  title={Robust Speech Recognition via Large-Scale Weak Supervision},
  author={Radford, Alec and Kim, Jong Wook and Xu, Tao and Brockman, Greg and McLeavey, Christine and Sutskever, Ilya},
  booktitle={International Conference on Machine Learning (ICML)}, pages={28492--28518}, year={2023}}

@article{liu2024kan,
  title={{KAN}: Kolmogorov-Arnold Networks},
  author={Liu, Ziming and Wang, Yixuan and Vaidya, Sachin and Ruehle, Fabian and Halverson, James and Soljacic, Marin and Hou, Thomas Y. and Tegmark, Max},
  journal={arXiv preprint arXiv:2404.19756}, year={2024}}

@article{desplanques2020ecapa,
  title={{ECAPA-TDNN}: Emphasized Channel Attention, Propagation and Aggregation in {TDNN} Based Speaker Verification},
  author={Desplanques, Brecht and Thienpondt, Jenthe and Demuynck, Kris},
  journal={arXiv preprint arXiv:2005.07143}, year={2020}}

@inproceedings{ganin2015dann,
  title={Unsupervised Domain Adaptation by Backpropagation},
  author={Ganin, Yaroslav and Lempitsky, Victor},
  booktitle={International Conference on Machine Learning (ICML)}, year={2015}}

@inproceedings{khosla2020supcon,
  title={Supervised Contrastive Learning},
  author={Khosla, Prannay and Teterwak, Piotr and Wang, Chen and Sarna, Aaron and Tian, Yonglong and Isola, Phillip and Maschinot, Aaron and Liu, Ce and Krishnan, Dilip},
  booktitle={Advances in Neural Information Processing Systems (NeurIPS)}, year={2020}}

@inproceedings{park2019specaugment,
  title={{SpecAugment}: A Simple Data Augmentation Method for Automatic Speech Recognition},
  author={Park, Daniel S. and Chan, William and Zhang, Yu and Chiu, Chung-Cheng and Zoph, Barret and Cubuk, Ekin D. and Le, Quoc V.},
  booktitle={Interspeech}, year={2019}}

@inproceedings{dai2021coatnet,
  title={{CoAtNet}: Marrying Convolution and Attention for All Data Sizes},
  author={Dai, Zihang and Liu, Hanxiao and Le, Quoc V. and Tan, Mingxing},
  booktitle={Advances in Neural Information Processing Systems (NeurIPS)}, year={2021}}

@misc{oshea2015cnn,
  title={An Introduction to Convolutional Neural Networks},
  author={O'Shea, Keiron and Nash, Ryan}, year={2015}, note={arXiv:1511.08458}}

@article{trockman2022convmixer,
  title={Patches Are All You Need?},
  author={Trockman, Asher and Kolter, J. Zico},
  journal={arXiv preprint arXiv:2201.09792}, year={2022}}

@inproceedings{maurya2026decker,
  title={{DECKER}: Domain-invariant Embedding for Cross-Keyboard Extraction and Recognition},
  author={Maurya, Bikrant Bikram Pratap and Choudhury, Nitin and Agarwal, Daksh and Buduru, Arun Balaji},
  booktitle={Proceedings of the ACM Asia Conference on Computer and Communications Security (ASIACCS)},
  pages={1707--1720}, year={2026},
  doi={10.1145/3779208.3805984},
  note={\url{https://dl.acm.org/doi/abs/10.1145/3779208.3805984}}}

@article{choudhury2026keyac,
  title={Impact Analysis of Speech Representation Learning Models for Acoustic Side-Channel Attack},
  author={Choudhury, Nitin and Maurya, Bikrant Bikram Pratap and Buduru, Arun Balaji and Phukan, Orchid Chetia},
  journal={arXiv preprint arXiv:2606.21210}, year={2026},
  note={\url{https://arxiv.org/abs/2606.21210}}}

@inproceedings{maurya2026evflow,
  title={Surviving the Codec: Continuous Sparse Transduction for Acoustic Side-Channel Attacks over {VoIP}},
  author={Maurya, Bikrant Bikram Pratap and Choudhury, Nitin and Buduru, Arun Balaji},
  booktitle={Interspeech (under review)}, year={2026}}

@article{achiam2023gpt4,
  title={{GPT-4} Technical Report},
  author={Achiam, Josh and others}, journal={arXiv preprint arXiv:2303.08774}, year={2023}}

@article{touvron2023llama,
  title={{LLaMA}: Open and Efficient Foundation Language Models},
  author={Touvron, Hugo and others}, journal={arXiv preprint arXiv:2302.13971}, year={2023}}

@inproceedings{brown2020gpt3,
  title={Language Models are Few-Shot Learners},
  author={Brown, Tom B. and Mann, Benjamin and Ryder, Nick and Subbiah, Melanie and Kaplan, Jared and Dhariwal, Prafulla and Neelakantan, Arvind and Shyam, Pranav and Sastry, Girish and Askell, Amanda and others},
  booktitle={Advances in Neural Information Processing Systems (NeurIPS)}, year={2020}}

@inproceedings{liu2022bidirectional,
  title={Bi-Directional Modality Fusion Network for Audio-Visual Event Localization},
  author={Liu, Shuo and Quan, Weize and Liu, Yuan and Yan, Dong-Ming},
  booktitle={IEEE International Conference on Acoustics, Speech and Signal Processing (ICASSP)},
  pages={4868--4872}, year={2022}, doi={10.1109/ICASSP43922.2022.9746280}}

@article{phukan2023transforming,
  title={Transforming the Embeddings: A Lightweight Technique for Speech Emotion Recognition Tasks},
  author={Phukan, Orchid Chetia and Buduru, Arun Balaji and Sharma, Rajesh},
  journal={arXiv preprint arXiv:2305.18640}, year={2023}}

@article{yang2021superb,
  title={{SUPERB}: Speech Processing Universal PERformance Benchmark},
  author={Yang, Shu-wen and Chi, Po-Han and Chuang, Yung-Sung and Lai, Cheng-I Jeff and Lakhotia, Kushal and Lin, Yist Y. and Liu, Andy T. and Shi, Jiatong and Chang, Xuankai and Lin, Guan-Ting and others},
  journal={arXiv preprint arXiv:2105.01051}, year={2021}}

@inproceedings{shon2022slue,
  title={{SLUE}: New Benchmark Tasks for Spoken Language Understanding Evaluation on Natural Speech},
  author={Shon, Suwon and Pasad, Ankita and Wu, Felix and Brusco, Pablo and Artzi, Yoav and Livescu, Karen and Han, Kyu J.},
  booktitle={IEEE International Conference on Acoustics, Speech and Signal Processing (ICASSP)},
  pages={7927--7931}, year={2022}}

@inproceedings{shon2023slue2,
  title={{SLUE} Phase-2: A Benchmark Suite of Diverse Spoken Language Understanding Tasks},
  author={Shon, Suwon and Arora, Siddhant and Lin, Chyi-Jiunn and Pasad, Ankita and Wu, Felix and Sharma, Roshan and Wu, Wei-Lun and Lee, Hung-yi and Livescu, Karen and Watanabe, Shinji},
  booktitle={Annual Meeting of the Association for Computational Linguistics (ACL)}, pages={8906--8937}, year={2023}}

@inproceedings{kodali2023,
  title={Classification of Vocal Intensity Category from Speech using the wav2vec2 and Whisper Embeddings},
  author={Kodali, Manila and Kadiri, Sudarsana and Alku, Paavo},
  booktitle={Interspeech}, pages={4134--4138}, year={2023}}

@article{yang2024largescale,
  title={A Large-Scale Evaluation of Speech Foundation Models},
  author={Yang, Shu-wen and Chang, Heng-Jui and Huang, Zili and Liu, Andy T. and Lai, Cheng-I and Wu, Haibin and Shi, Jiatong and Chang, Xuankai and Tsai, Hsiang-Sheng and Huang, Wen-Chin and others},
  journal={IEEE/ACM Transactions on Audio, Speech, and Language Processing}, volume={32}, pages={2884--2899}, year={2024}}

@article{porjazovski2024,
  title={From Raw Speech to Fixed Representations: A Comprehensive Evaluation of Speech Embedding Techniques},
  author={Porjazovski, Dejan and Gr{\'o}sz, Tam{\'a}s and Kurimo, Mikko},
  journal={IEEE/ACM Transactions on Audio, Speech, and Language Processing}, volume={32}, pages={3546--3560}, year={2024}}

@inproceedings{novotny2018xvectors,
  title={On the Use of X-vectors for Robust Speaker Recognition},
  author={Novotn{\`y}, Ond{\v{r}}ej and Plchot, Old{\v{r}}ich and Matejka, Pavel and Mosner, Ladislav and Glembek, Old{\v{r}}ich},
  booktitle={Odyssey: The Speaker and Language Recognition Workshop}, pages={168--175}, year={2018}}

@inproceedings{cai2025loki,
  title={{LoKI}: Low-dimensional {KAN} for Efficient Fine-tuning Image Models},
  author={Cai, Xin and Pan, Rui and Yang, Hui},
  booktitle={Computer Vision and Pattern Recognition Conference (CVPR)}, pages={14869--14880}, year={2025}}

@article{li2024grkan,
  title={From {KAN} to {GR-KAN}: Advancing Speech Enhancement with KAN-based Methodology},
  author={Li, Haoyang and Hu, Yuchen and Chen, Chen and Siniscalchi, Sabato Marco and Liu, Shixiong and Chng, Eng Siong},
  journal={arXiv preprint arXiv:2412.17778}, year={2024}}

@inproceedings{phukan2025rethinking,
  title={Rethinking Cross-Corpus Speech Emotion Recognition Benchmarking: Are Paralinguistic Pre-trained Representations Sufficient?},
  author={Phukan, Orchid Chetia and Akhtar, Mohd Mujtaba and Behera, Swarup Ranjan and Bhagath, Pailla and Reddy, Priyabrata B. and Buduru, Arun Balaji and others},
  booktitle={Asia Pacific Signal and Information Processing Association Annual Summit and Conference (APSIPA ASC)}, pages={1022--1027}, year={2025}}

@article{choudhury2026foca,
  title={{FOCA}: Multimodal Malware Classification via Hyperbolic Cross-Attention},
  author={Choudhury, Nitin and Maurya, Bikrant Bikram Pratap and Phukan, Orchid Chetia and Buduru, Arun Balaji},
  journal={arXiv preprint arXiv:2601.17638}, year={2026}}

@article{choudhury2026gocoma,
  title={{GoCoMa}: Hyperbolic Multimodal Representation Fusion for Large Language Model-Generated Code Attribution},
  author={Choudhury, Nitin and Maurya, Bikrant Bikram Pratap and Kuwar, Bhuvnesh V. and Buduru, Arun Balaji},
  journal={arXiv preprint arXiv:2604.16377}, year={2026}}

@article{gong2021ast,
  title={{AST}: Audio Spectrogram Transformer},
  author={Gong, Yuan and Chung, Yu-An and Glass, James},
  journal={arXiv preprint arXiv:2104.01778}, year={2021}}

@inproceedings{liu2025flowmatching,
  title={Flow Matching for Collaborative Filtering},
  author={Liu, Chengkai and Zhang, Yangtian and Wang, Jianling and Ying, Rex and Caverlee, James},
  booktitle={ACM SIGKDD Conference on Knowledge Discovery and Data Mining (KDD)}, pages={1765--1775}, year={2025}}

@article{gu2024mamba,
  title={Mamba: Linear-Time Sequence Modeling with Selective State Spaces},
  author={Gu, Albert and Dao, Tri},
  journal={arXiv preprint arXiv:2312.00752}, year={2024}}

@inproceedings{graves2006ctc,
  title={Connectionist Temporal Classification: Labelling Unsegmented Sequence Data with Recurrent Neural Networks},
  author={Graves, Alex and Fern{\'a}ndez, Santiago and Gomez, Faustino and Schmidhuber, J{\"u}rgen},
  booktitle={International Conference on Machine Learning (ICML)}, pages={369--376}, year={2006}}

@incollection{cover2006infotheory,
  title={Entropy, Relative Entropy and Mutual Information},
  author={Cover, Thomas M. and Thomas, Joy A.},
  booktitle={Elements of Information Theory}, publisher={John Wiley \& Sons}, pages={12--49}, year={2006}}

\clearpage
\chapter*{List of Publications}
\addcontentsline{toc}{chapter}{List of Publications}
\markboth{List of Publications}{}

The following peer-reviewed publications form the basis of this thesis. All studies use the
\hear{} dataset introduced in Chapter~\ref{ch:hear}, publicly documented at
\url{https://github.com/vikrant-vikram/Hear-Dataset}.

\vspace{0.4cm}
\noindent\textbf{Conference Publications}
\begin{enumerate}[leftmargin=1.6em,itemsep=6pt]

\item Bikrant Bikram Pratap Maurya, Nitin Choudhury, Daksh Agarwal, and Arun Balaji Buduru.
``DECKER: Domain-invariant Embedding for Cross-Keyboard Extraction and Recognition.''
In \emph{Proceedings of the ACM Asia Conference on Computer and Communications Security (ASIACCS)}, Bangalore, India, 2026, pp.\ 1707--1720.
\newline\url{https://dl.acm.org/doi/abs/10.1145/3779208.3805984}
\newline\emph{(Core of the thesis: Chapters~\ref{ch:hear}, \ref{ch:decker}, and \ref{ch:results}.)} \cite{maurya2026decker}

\item Nitin Choudhury, Bikrant Bikram Pratap Maurya, Arun Balaji Buduru, and Orchid Chetia Phukan.
``Impact Analysis of Speech Representation Learning Models for Acoustic Side-Channel Attack.''
In \emph{Proceedings of Interspeech}, 2026.
\newline\url{https://arxiv.org/abs/2606.21210}\cite{choudhury2026keyac}

\item Nitin Choudhury, Bikrant Bikram Pratap Maurya, Orchid Chetia Phukan, and Arun Balaji Buduru.
``FOCA: Multimodal Malware Classification via Hyperbolic Cross-Attention.''
In \emph{Proceedings of IEEE ICASSP}, 2026.
\newline\url{https://arxiv.org/abs/2601.17638}\cite{choudhury2026foca}

\item Nitin Choudhury, Bikrant Bikram Pratap Maurya, Bhavinkumar Vinodbhai Kuwar, and Arun Balaji Buduru.
``GoCoMA: Hyperbolic Multimodal Representation Fusion for Large Language Model-Generated Code Attribution.''
In \emph{Proceedings of IEEE ICME}, 2026.
\newline\url{https://arxiv.org/abs/2604.16377}\cite{choudhury2026gocoma}

\item
``CORAL: A Synergistic Approach to Deception Detection with Socially-Aware Representations and Positive-Unlabeled Learning.''
In \emph{Proceedings of AACL-IJCNLP 2025}, Student Research Workshop.

\end{enumerate}

\vspace{0.2cm}
\noindent\footnotesize{* Equal contribution.}

\clearpage

\end{document}